\documentstyle[12pt,axodraw]{article}
\catcode`@=11
\newcount\@tempcntc
\def\@citex[#1]#2{\if@filesw\immediate\write\@auxout{\string\citation{#2}}\fi
  \@tempcnta\z@\@tempcntb\m@ne\def\@citea{}\@cite{\@for\@citeb:=#2\do
    {\@ifundefined
       {b@\@citeb}{\@citeo\@tempcntb\m@ne\@citea\def\@citea{,}{\bf ?}\@warning
       {Citation `\@citeb' on page \thepage \space undefined}}%
    {\setbox\z@\hbox{\global\@tempcntc0\csname b@\@citeb\endcsname\relax}%
     \ifnum\@tempcntc=\z@ \@citeo\@tempcntb\m@ne
       \@citea\def\@citea{,}\hbox{\csname b@\@citeb\endcsname}%
     \else
      \advance\@tempcntb\@ne
      \ifnum\@tempcntb=\@tempcntc
      \else\advance\@tempcntb\m@ne\@citeo
      \@tempcnta\@tempcntc\@tempcntb\@tempcntc\fi\fi}}\@citeo}{#1}}
\def\@citeo{\ifnum\@tempcnta>\@tempcntb\else\@citea\def\@citea{,}%
  \ifnum\@tempcnta=\@tempcntb\the\@tempcnta\else
   {\advance\@tempcnta\@ne\ifnum\@tempcnta=\@tempcntb \else \def\@citea{--}\fi
    \advance\@tempcnta\m@ne\the\@tempcnta\@citea\the\@tempcntb}\fi\fi}
\catcode`@=12

\begin{document}

\begin{flushright}
RAL-TR/96--029\\
May 1996
\end{flushright}

\begin{center}
{\LARGE {\bf Gauge-Invariant Resummation Formalism }}\\[0.4cm]
{\LARGE {\bf for Two-Point Correlation Functions }}\\[2.4cm]
{\large Joannis Papavassiliou}$^a$\footnote[1]{e-mail address: 
papavass@cpt.univ-mrs.fr}
{\large and Apostolos Pilaftsis}$^b$\footnote[2]{E-mail address: 
pilaftsis@v2.rl.ac.uk}\footnotetext{Address after 1st of October: 
Max-Planck-Institute, F\"ohringer Ring 6, D-80805 M\"unchen, FRG.}\\[0.4cm]
$^a${\em Centre de Physique Th\'eorique, CNRS, Luminy,}\\ 
{\em F-13288 Marseille CEDEX 09, France}\\[0.3cm]
$^b${\em Rutherford Appleton Laboratory, Chilton, Didcot, Oxon, OX11 0QX, UK}
\end{center}
\vskip1.5cm
\centerline{\bf ABSTRACT}
The consistent description of unstable particles, renormalons, or other
Schwinger--Dyson-type of solutions within the framework of perturbative gauge
field theories necessitates the definition and resummation of off-shell Green's
functions, which must respect several crucial physical requirements.  A
formalism is presented for resummation of off-shell two-point correlation
functions, which is mainly based on arguments of analyticity, unitarity, gauge
invariance  and renormalizability. The  analytic results obtained with various
methods, including the background field gauges and  the  pinch technique are
confronted with the physical requirements imposed; to one-loop order the pinch
technique approach satisfies all of them. Using renormalization group
arguments, we discuss issues  of uniqueness of the resummation procedure
related to the latter method.\\[0.3cm] 
PACS nos.: 14.70.Fm, 11.15.Bt, 11.15.Ex

\newpage

\section{Introduction}
\indent

It is well known that in non-Abelian gauge theories individual off-shell
Green's functions are in general plagued with various pathologies, such as
gauge dependences, bad high energy behaviour, or lack of renormalizability,
which, strictly speaking, render them void of any physical meaning. To the
extend that the physical issues at hand can be dealt with within the confines
of conventional perturbation theory, the aforementioned pathologies pose no
real problem. Indeed, when combined together to form observables, the
individually pathological Green's functions conspire in such a way as to give
a physically meaningful answer, order by order in perturbation theory. A
classic example of the subtle cancellation mechanisms in effect is the
computation of electroweak $S$-matrix elements in the unitary gauge; there,
even though the conventional two-, three- and four- point functions are not
even renormalizable, the final $S$-matrix element turns out to be
well-defined. 

There is, however, a plethora of physically important questions, which cannot be
treated in the framework of conventional perturbation theory. In quantum
chromodynamics (QCD) for example, the only known way to study in the continuum
phenomena, such as chiral symmetry breaking or gluon mass generation, is by
means of the Schwinger-Dyson equations \cite{JMC}. Here, the pathologies of the
Green's functions start playing a r\^ole. Indeed, the Schwinger-Dyson equation
are build up by off-shell Green's functions; if one could solve these equations
exactly, the Green's functions obtained would again conspire to yield
physically meaningful answers. However, since the Schwinger-Dyson series
constitutes an infinite set of  coupled non-linear integral equations, a
truncation is necessary, which, if carried out casually, may give rise to
physically meaningless answers, such as gauge-dependent expressions for
ostensibly gauge independent, physical quantities.

Even though the need for a self-consistent scheme for constructing off-shell
Green's functions is more or less expected when dealing with a strongly
coupled theory such as QCD, perhaps the most compelling physical circumstances
advocating its necessity have been encountered in the context of a ``weakly''
coupled theory, namely the electroweak $SU(2)_L\otimes U(1)_Y$ model
\cite{AP,AS,JP&AP}. Indeed, the presence of unstable particles makes it
impossible to compute physical amplitudes for arbitrary values of the
kinematic parameters, unless a resummation has first taken place. Simply
stated, perturbation theory breaks down in the vicinity of resonances, and
information about the dynamics to ``all orders'' needs be encoded already at
the level of Born amplitudes. As was already  pointed out in \cite{AP}, if one
attempts to naively promote Veltman's formalism for scalar theories
\cite{veltman} to the case of gauge theories, one is invariably led to gross
violations of gauge invariance and unitarity. As explained in \cite{JP&AP},
resumming the conventional two-point function of a gauge boson in order to
construct a Breit-Wigner type of propagator, takes into account higher order
corrections for only certain parts of the Born amplitude, whereas crucial
contributions originating from box and vertex graphs are not included
properly. As a result, the subtle cancellation mechanism alluded to before,
even though in reality is still in effect, gets distorted by the casual
resummation, resulting in artifacts, which thwart the predictive power of
$S$-matrix perturbation theory. 

Given the subtle nature of the problem, the question naturally arises, what
set of physical criteria must be satisfied by a resummation algorithm, in
order for it to qualify as ``physical''. In other words, what are the guiding
principles, which will allow one to determine whether or not the resummed
quantity carries any physically meaningful information, and to what extend it
captures the essential underlying dynamics? To address these questions in
this paper, we postulate a set of field-theoretical requirements that we
consider crucial when attempting to define a proper resummed propagator. Our
considerations propose an answer to the question of how to analytically
continue the Lehmann--Symanzik--Zimmermann (LSZ) formalism~\cite{LSZ} in the
off-shell region of Green's functions in a way which is manifestly
gauge-invariant and consistent with unitarity. In addition, we demonstrate
that the off-shell Green's functions obtained by the Pinch Technique (PT)
\cite{PT,JPself,PTall,NJW} satisfy all these requirements. In fact, these
requirements are, in a way, inherent within the PT approach, as we will see in
detail in what follows. 

In particular, the following is required from an off-shell, one-particle
irreducible (1PI), effective  two-point function:
\begin{itemize}

\item[(i)] {\em Resummability}. The effective two-point functions must be
resummable. For the conventionally defined two-point functions, the
resummability can be formally derived from the path integral. In the $S$-matrix
PT approach, the resummability of the effective two-point functions is more
involved and must be based on a careful analysis of the structure of the
$S$-matrix to higher orders in perturbation theory \cite{JP&AP}. 

\item[(ii)] {\em Analyticity of the off-shell Green's function}. An analytic
two-point function has the property that its real and imaginary parts are
related by a dispersion relation (DR), up to a maximum number of two
subtractions. The latter is a necessary condition when considering
renormalizable Green's functions, as we will discuss in Section 2. 

\item[(iii)] {\em Unitarity and the optical relation}. In the conventional
framework, unitarity is defined only for on-shell $S$-matrix elements, leading
to the familiar optical theorem (OT) for the forward scattering. Here, we
postulate the validity of the optical relation for the off-shell Green's
function, when embedded in an $S$-matrix element, in a way which will become
clear in what follows. An important consequence of this requirement is that
the imaginary part of the off-shell Green's function should not contain any
unphysical thresholds. As a counter-example, in Section 7, it will be shown
that this pathology is in fact induced by the quantum fields in the 
background-field-gauge (BFG) method \cite{BFG} for $\xi_Q\not= 1$. 

\item[(iv)] {\em Gauge invariance}. As has been mentioned above, one has to
require that the effective Green's functions are gauge-fixing parameter (GFP)
independent and satisfy WIs in compliance with the classical action. For
instance, the latter is guaranteed in the BFG method but not the former. This
condition also guarantees that gauge invariance does not get spoiled after
Dyson summation of the GFP-independent self-energies. In some of the recent
literature, the terms of gauge invariance and gauge independence have been used
for two different aspects. For example, in the BFG the classical background
fields respect gauge invariance in the classical action. However, this fact
does not ensure that the quantum fields respect some form of quantum gauge
invariance, neither does imply that some kind of a Becchi-Rouet-Stora (BRS)
symmetry \cite{BRS} is present for the fields inside the quantum loops after
fixing the gauge of the theory \cite{Morris,Schenk}. In our discussion, when
referring to gauge invariance, we will encompass both meanings, {\em i.e.},
gauge invariance of the tree-level classical particles as well as BRS
invariance of the quantum fields. A direct but non-trivial consequence of the
gauge invariance and of the abelian-type WIs that the effective off-shell
Green's functions satisfy is that for large asymptotic momenta transfers
($s\to\infty$), the self-energy under construction must capture the running of
the gauge coupling, as it happens in quantum electrodynamics (QED). Because of
the abelian-type WIs and on account of resummation, the above argument can be
generalized to $n$-point functions. In addition, the off-shell $n$-point
transition amplitudes should display the correct high-energy limit as is
dictated by the Equivalence Theorem \cite{EqTh}.

\item[(v)] {\em Multiplicative renormalization}. Since we are interested in
renormalizable theories, {\em i.e.}, theories containing operators of
dimension no higher than four, the off-shell Green's functions calculated
within an approach should admit renormalization. However, this requirement
alone is not sufficient when resummation is considered. The appearance of a
two-point function in the denominator of a resummed propagator makes it
unavoidable to demand that renormalization be {\em multiplicative}; otherwise,
the analytic expressions will suffer from spurious ultraviolet (UV)
divergences. Particular examples of the kind are some ghost-free gauges, such
as the light-cone or planar gauge \cite{planar}. 

\item [(vi)] {\em Position of the pole}. Since the position of the pole is the
only gauge-invariant quantity that one can extract from conventional
self-energies, any acceptable resummation procedure should give rise to
effective self-energies which do not shift the position of the pole. This
requirement drastically reduces the arbitrariness in constructing effective
two-point correlation function.

\end{itemize}

A closer look at these requirements reveals that they are in fact very tightly
interwoven; relaxing even one of them could give rise to unphysical results,
sometimes in rather subtle ways. As an example of the subtleties involved, we
investigate the BFG \cite{BFG,DDW} in Section 8. Despite the fact that the
background fields of the BFG obey the Ward identities (WIs) of the classical
Lagrangian, even after quantizing the theory, the BFG expressions for the
self-energies depend explicitly on the quantum gauge parameter $\xi_Q$; in
turn, in theories with spontaneous symmetry breaking (SSB), this dependence on
$\xi_{Q}$ gives rise to {\em unphysical} threshold channels for $\xi_Q\not=
1$. Obviously, such unphysical absorptive contributions should not be resummed
to all orders. In fact, we find that the sub-amplitudes containing physical
Landau singularities and those, which do not, satisfy the same BFG WIs. Only
the case of BFG with $\xi_Q=1$ is free from unphysical poles, and the results
of the Green's functions collapse to these of the PT. Evidently, relaxing the
requirement of GFP independence, by allowing $\xi_{Q}$ to survive, interferes
with unitarity in a non-trivial way. 

We now present a roadmap of our paper: In Section 2, we review the crucial
properties of analyticity of two-point correlation functions. We then derive
some important consequences arising from DRs, which should be satisfied by a
consistent analytic approach. The results of this analysis may also be applied
to eliminate a large degree of arbitrariness in defining off-shell transition
amplitudes. Issues of renormalization are also discussed. 

In Section 3, we discuss the r\^ole of unitarity and OT and elucidate its
connection with gauge invariance. In Section 4, we show how to employ
unitarity, analyticity and elementary tree-level WIs (EWIs), in order to
obtain a self-consistent picture in the context of QCD. In particular, we work
with the right hand side (RHS) of the OT, where only physical particles (no
ghosts) appear as intermediate states. In Section 5, we focus again on the
same process as in the previous section and present a different (equivalent
but non-trivial) point of view. In particular, we start again from the RHS of
the OT and show how the unitarity of an on-shell transition amplitude and the
BRS symmetry \cite{BRS} of the quantum action can be exploited to reinforce
gauge invariance and GFP independence for off-shell Green's
functions. In the context of one-loop QCD, these properties rigorously prove
the independence of the PT on the gauge-fixing procedure. 

In Section 6, the analysis of Section 5 is extended to the case of the minimal
Standard Model (SM). We concentrate on a charged process with {\em
non-conserved} external currents and resort again to the (slightly more
involved) EWIs. The propagator-like expression obtained by working with the
RHS of the OT is then fed into a twice subtracted DR.  The result obtained is
identical to the real part of the PT $W$-boson self-energy, already known from
previous considerations. This example convincingly demonstrates the combined
power of unitarity and analyticity. In Section 7, we take a different point of
view and work directly with the left-hand-side (LHS) of the OT, where
``unphysical'' degrees of freedom, such as ghosts and would-be Goldstone
bosons, appear now as intermediate states. Using the usual Cutkosky rules, and
exploiting again the EWIs of the theory to the fullest, we arrive at the
imaginary part of the PT $W$-boson self-energy. This constitutes a highly
non-trivial self-consistency check, demonstrating that as long as one fully
exploits the elementary symmetries of the theory, one can work freely with
either side of the optical relation, arriving at the same physically
consistent results. 

In Section 8, we turn our attention to the BFG and show that the dependence of
the resummed BFG two-point functions on the ``quantum'' GFP $\xi_{Q}$ is far
from innocuous, leading to the violation of unitarity, because of the
appearance of unphysical thresholds. Furthermore, the physical and unphysical
expressions are found to satisfy exactly the same tree-level WIs. This fact
demonstrates beyond any doubt that a combination of requirements need be
imposed in order to arrive at a physically reliable result. Indeed, satisfying
external tree-level WIs is a necessary but not sufficient requirement in this
context.

In Section 9, we show under mild assumptions that the PT resummation gives
rise to ``unique'' results \cite{RGE}. By ``unique'', we mean that at the end
of the PT rearrangement, and after renormalization has been completed,  no
further pieces may be moved around without leading to a violation of some of
the physical properties characterizing the PT Green's functions. Finally, we
present our conclusions in Section 10.

\setcounter{equation}{0}
\section{Analyticity and renormalization}
\indent

Analyticity is one of the most important properties that governs physical
transition amplitudes. Correlation functions are considered to be analytic in
their kinematic variables, which is expressed by means of the so-called
DRs~\cite{KK,DR,ELOP}. In this section, we briefly review some important facts
about DRs and renormalization and discuss the subtleties encountered in
non-Abelian gauge theories. 

If a complex function $f(z)$ is analytic in the interior of and upon a closed
curve, $C_\uparrow$ say in Fig.\ 1, and $x+i\varepsilon$ (with $x,\varepsilon
\in$ {\bf R} and $\varepsilon >0$) is a point within the closed curve
$C_\uparrow$, we then have the Cauchy's integral form, 
\begin{equation}
f(x+i\varepsilon)\ =\ \frac{1}{2\pi i} \oint_{C_\uparrow} dz\,
\frac{f(z)}{z-x-i\varepsilon}\ ,
\end{equation} 
where $\oint$ denotes that the path $C_\uparrow$ is singly wound. Using
Schwartz's reflection principle, one also obtains
\begin{equation}
f(x-i\varepsilon)\ =\ -\, \frac{1}{2\pi i} \oint_{C_\downarrow} dz\,
\frac{f(z)}{z-x+i\varepsilon}\ .
\end{equation} 
Note that $C_\uparrow^*=C_\downarrow$. Sometimes, an analytic function is 
called holomorphic; both terms are equivalent for complex functions.
\begin{center}
\begin{picture}(300,200)(0,0)
\SetWidth{0.8}
\Line(100,105)(200,105)\ArrowLine(130,105)(150,105)
\CArc(150,105)(50,0,180)\ArrowArc(150,105)(50,30,60)
\Line(100,95)(200,95)\ArrowLine(130,95)(150,95)
\CArc(150,95)(50,180,360)\ArrowArcn(150,95)(50,330,300)
\LongArrow(80,100)(220,100)\LongArrow(150,30)(150,170)
\Text(215,92)[l]{$\Re e z$}\Text(145,170)[r]{$\Im m z$}
\Text(175,110)[]{$\bullet $} \Text(175,90)[]{$\bullet $}
\Text(177,120)[]{$x+i\varepsilon$}\Text(177,82)[]{$x-i\varepsilon$}
\DashLine(150,100)(120,145.3){2.} \DashLine(150,100)(120,54.8){2.}
\Text(130,125)[r]{$R$}\Text(130,75)[r]{$R$}
\ArrowArc(150,100)(10,0,122.)\Text(152,116)[l]{$\theta$}
\Text(185,147)[l]{$C_\uparrow$}\Text(185,51)[l]{$C_\downarrow$}
\end{picture}\\
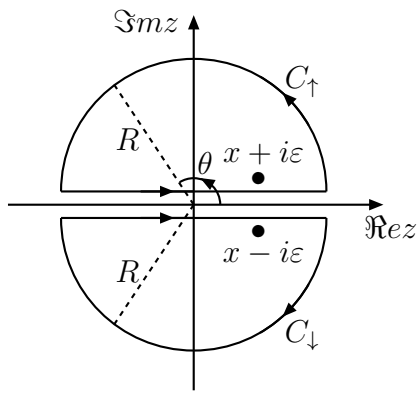
{\bf Fig. 1:} Contours of complex integration 
\end{center}

Of significant importance in the discussion of physical processes is a DR,
which relates the imaginary part of an analytic function $f(x)$ to its real
part, and vice versa. We assume for the moment that the analytic function
$f(z)$ has the asymptotic behaviour, $|f(z)|\le C/R^k$, for large radii $R$ as
shown in Fig.\ 1, where  $C$ is a real nonnegative constant and $k>0$; this
assumption will be relaxed later on, giving rise to more involved DR.  Taking
now the limit $\varepsilon\to 0$, it is easy to evaluate $\Re e f(x)$ through 
\begin{equation} 
2\Re e f(x)\ =\ `\lim_{\varepsilon\to 0}\mbox{'}\Big[ f(x+i\varepsilon)
+ f^*(x-i\varepsilon)\Big]\ =\ `\lim_{\varepsilon\to 0}\mbox{'}\, 
\frac{1}{\pi}\int\limits_{-\infty}^{+\infty}dx'\, 
\Im m \left( \frac{f(x')}{x'-x-i\varepsilon}\right)\, +\, \Gamma_{\infty}.
\end{equation}
Here, $`\lim_{\varepsilon\to 0}\mbox{'}$ means that the limit
should be taken {\em after} the integration has been performed, and 
\begin{equation}
\label{orio}
\Gamma_\infty\ =\ \frac{1}{\pi}\lim_{R\to \infty}\, \Re e\,
\int_0^\pi d\theta\, f(Re^{i\theta})\, .
\end{equation}
Because of the assumed asymptotic behaviour of $f(z)$ at infinity, the
integral over the upper infinite semicircle in Fig.\ 1, $\Gamma_{\infty}$, can
be easily shown to vanish. Employing the well-known identity for
distributions, 
\begin{displaymath}
`\lim_{\varepsilon\to 0}\mbox{'}\, \frac{1}{x'-x-i\varepsilon}\ =\
\mbox{P}\frac{1}{x'-x}\ +\ i\pi\delta (x'-x),
\end{displaymath}
we arrive at the unsubtracted DR,
\begin{equation}
\label{DR1}
\Re e f(x)\ =\ \frac{1}{\pi}\, \mbox{P}\, \int\limits^{+\infty}_{-\infty}
dx'\, \frac{\Im m f(x')}{x'-x}\ . 
\end{equation} 
In Eq.\ (\ref{DR1}), the symbol P in front of the integral stands for
principle value integration. Following a similar line of arguments, one can
express the imaginary part of $f(x)$ as an integral over $\Re e f(x)$. 

In the previous derivation, the assumption that $|f(z)|$ approaches zero
sufficiently fast at infinity has been crucial, since it guarantees that
$\Gamma_{\infty}\to 0$. However, if we were to relax this assumption,
additional subtractions need be included in order to arrive at a finite
expression. For instance, for $|f(z)|\le CR^k$ with $k<1$, it is sufficient to
carry out a single subtraction at a point $x=a$. In this way, one has 
\begin{equation}
\label{DR2}
\Re e f(x)\ =\ \Re e f(a)\ +\ \frac{(x-a)}{\pi}\, \mbox{P}\, 
\int\limits_{-\infty}^{+\infty}dx'\, \frac{\Im m f(x')}{(x'-a)(x'-x)}\ .
\end{equation}
${}$From Eq.\ (\ref{DR2}), it is obvious that $\Re e f(x)$ can entirely be
obtained from $\Im m f(x)$, up to a unknown, real constant $\Re e f(a)$.
Usually, the point $a$ is chosen in a way such that $\Re e f(a)$ takes a
specific value on account of some physical requirement.  For example, if $\Im
m f(q^2)$ is the imaginary part of the magnetic form factor of an electron
with photon virtuality $q^2$, one can prescribe that the physical condition
$\Re e f(0)=0$ should hold true in the Thomson limit. 

We next focus on the study of some crucial analytic properties of off-shell
transition amplitudes within the context of renormalizable field theories. In
such theories, one is allowed to have at most two subtractions for a two-point
correlation function. If $\Pi (s)$ is the self-energy function of a scalar
particle with mass $m$ and off-shell momentum $q$ ($s=q^2$) ---the fermionic
or vector case is analogous---  then the real (or dispersive) part of this
amplitude can be fully determined by its imaginary (or absorptive) part via
the expression 
\begin{equation}
\label{DR3}
\Re e \Pi (s)\ =\ \Re e \Pi (m^2)\, +\, (s-m^2)\Re e \Pi'(m^2)\, +\, 
\frac{(s-m^2)^2}{\pi}\mbox{P}\, \int\limits_{0}^{+\infty} ds'\, 
\frac{\Im m \Pi (s')}{(s'-m^2)^2(s'-s)}\ .\quad
\end{equation}
${}$From Eq.\ (\ref{DR3}), one can readily see that the two subtractions, $\Re
e \Pi (m^2)$ and the derivative $\Re e \Pi'(m^2)$, correspond to the mass and
wave-function renormalization constants in the on-mass shell (OS) scheme,
respectively. At higher orders, internal renormalizations of $\Im m \Pi (s)$,
due to counterterms (CTs) coming from lower orders, should also be taken into
account. Then, Eq.\ (\ref{DR3}) is still valid, {\em i.e.}, it holds to order
$n$ provided $\Im m \Pi (s)$ is renormalized to order $n-1$. In general, the
function $\Im m \Pi (s)$ has its support in the non-negative real axis, {\em
i.e.}, for $s\ge 0$. This can be attributed to the semi-boundness of the
spectrum of the Hamiltonian, Spec$H\ge 0$~\cite{Khalfin}. Note that for
spectrally represented two-point correlation functions, we have the additional
condition  $\Im m \Pi( m^2 )\ge 0 ~$\cite{Bjorken,Justin}. 

As has been mentioned above, in renormalizable field theories it is required
that $\Pi (s)$ should be finite after two subtractions have been performed.
This implies that
\begin{equation} 
\label{asympt}
|\Pi (s) |\ \le\  C s^k\, ,\qquad \mbox{with}\quad k<2,
\end{equation} 
as $s\to \infty$. Obviously, the same inequality holds true for the real as
well as the imaginary part of $\Pi (s)$. In pure non-abelian Yang-Mills
theories, such as quark-less QCD, the transverse part, $\Pi_T(s)$, of the gluon
vacuum polarization behaves asymptotically as
\begin{displaymath}
\Pi_T (s)\ \to\ C\, s\Big(\ln \frac{s}{\mu^2}\Big)^n\ .
\end{displaymath}
This result is consistent with Eq.\ (\ref{asympt}), for any $n < \infty$.
Furthermore, we mention in passing that the Froissart--Martin bound
\cite{FroissartMartin}, 
\begin{equation}
\label{Frois}
|\Pi (s)|\ \leq\ C\, s^3\Big(\ln\frac{s}{s_0}\Big)^2,
\end{equation}
at $s\to \infty$, which may be derived from axiomatic methods of field theory
\cite{FMbound}, is weaker than Eq.\ (\ref{asympt}). The analytic expression of
gluon vacuum polarization satisfies Eq.\ (\ref{Frois}). As a counter-example
to this situation, we may consider the Higgs self-energy in the unitary gauge;
the absorptive part of the Higgs self-energy has an $s^2$ dependence at high
energies, and its resummation~\cite{VW} is therefore not justified. 

We will now illustrate how DRs work in practice in the context of a scalar
field theory. As an example, we consider a toy model with interaction
Lagrangian,
\begin{equation}
\label{Lint}
{\cal L}_{int}\ =\ \frac{\lambda}{2}\phi^2\Phi\, ,
\end{equation}
where $\lambda$ is a non-vanishing coupling constant of dimensions of mass. We
denote the mass of the scalar $\phi$ by $m$ and the one of the $\Phi$ by $M$
and assume that $M\ge m$. 
\begin{center}
\begin{picture}(300,100)(0,0)
\SetWidth{0.8}
\ArrowLine(90,50)(120,50)  \Text(100,60)[r]{$\Phi (p)$}
\ArrowLine(180,50)(210,50) \Text(200,60)[l]{$\Phi (p) $} 
\ArrowArcn(150,50)(30,180,360) \Text(150,90)[r]{$\phi$}
\ArrowArc(150,50)(30,180,360) \Text(150,10)[r]{$\phi$}
\DashLine(155,90)(155,10){2.} 
\end{picture}\\
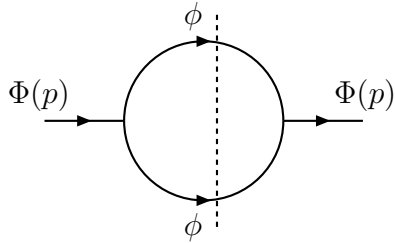
{\bf Fig. 2:} Two-point correlation function $\Pi_\Phi (s)$ at one loop 
\end{center}

One can calculate the imaginary part of the one-loop self-energy $\Pi_\Phi
(s)$ by using Cutkosky rules. The self-energy $\Pi_\Phi (s)$ develops a branch
cut for $s=p^2>4m^2$, which arises from the on-shell $\phi$-pair contribution
shown in Fig.\ 2. Thus, it is not difficult to obtain 
\begin{equation}
\label{ImPi}
\Im m \Pi_\Phi (s)\ =\ \frac{\lambda^2}{32\pi} 
\left( 1-\frac{4m^2}{s} \right)^{1/2}\, \theta (s-4m^2)\ .
\end{equation}
On the other hand, adopting dimensional regularization in dimensions
$D=4-2\epsilon $, we have
\begin{eqnarray}
\label{RePi}
\Pi_\Phi (s) &=& \frac{\lambda^2}{32\pi^2} \Bigg\{ \frac{1}{\epsilon}
- \gamma_E + \ln\frac{4\pi\mu^2}{m^2} + 2 - \Big( 1 - \frac{4m^2}{s} 
                             \Big)^{1/2}\nonumber\\
&&\times \ln\Bigg[
\frac{\displaystyle \Big( 1 - \frac{4m^2}{s} \Big)^{1/2} 
+ 1}{\displaystyle \Big( 1 - \frac{4m^2}{s} \Big)^{1/2} - 1}\Bigg]\Bigg\},
\end{eqnarray}
where $s$ should be analytically continued to $ s + i\varepsilon $. In fact,
for $s >4m^2$, the logarithmic function in Eq.\ (\ref{RePi}) assumes the form
\begin{displaymath}
\ln\Bigg[
\frac{\displaystyle 1 + \Big( 1 - \frac{4m^2}{s} \Big)^{1/2}}{\displaystyle 
1- \Big( 1 - \frac{4m^2}{s} \Big)^{1/2}}\Bigg]\ -\
i\pi\theta (s-4m^2)\, .
\end{displaymath}
Evidently, the absorptive part of $\Pi_\Phi (s)$ obtained from Eq.\
(\ref{RePi}) is equal to $\Im m \Pi_\Phi (s)$ in Eq.\ (\ref{ImPi}).
Furthermore, one can verify the validity of a DR of Eq. (\ref{DR2}), singly
subtracted at $s=0$. Since
\begin{equation}
\Re e \Pi_\Phi (0)\ =\ \frac{\lambda^2}{32\pi^2}\, \Big[ \frac{1}{\epsilon}
- \gamma_E + \ln\frac{4\pi\mu^2}{m^2} \Big],
\end{equation}
one can check that indeed,  
\begin{displaymath}
\frac{s}{\pi}\, \mbox{P} \int\limits_{4m^2}^{\infty}ds' 
\frac{\Im m \Pi_\Phi (s')}{s'(s'-s)}\ =\ \Re e\Pi_\Phi (s)\, -\, 
\Re e\Pi_\Phi (0)\, .
\end{displaymath}
This simple example explicitly demonstrates the analytic nature of a two-point
correlation function.

In the context of gauge field theories, one should anticipate a similar
analytic structure for two-point correlation functions. However, an extra
complication appears in such theories when off-shell transition amplitudes are
considered. In a theory with SSB, such as the SM for example, this
complication originates from the fact that, in addition to the physical
particles of the spectrum of the Hamiltonian, unphysical, gauge dependent
degrees of freedom, such as would-be Goldstone bosons and ghost fields make
their appearance. Although on-shell transition amplitudes contain only the
physical degrees of freedom of the particles involved on account of unitarity,
their continuation to the off-shell region is ambiguous, because of the
presence unphysical Landau poles, introduced by the aforementioned unphysical
particles. A reasonable prescription for accomplishing such an off-shell
continuation, which is very close in spirit to the previous example of the
scalar theory, would be to continue analytically an off-shell amplitude by
taking only {\em physical} Landau singularities into account. 

Consider for example the off-shell propagator of a gauge particle in the
conventional $R_\xi$ gauges or BFGs, which runs inside a quantum loop, {\em
viz.} 
\begin{equation}
\label{Prop1}
\Delta_{0\mu\nu}^{(\xi_Q)}(q)\ =\ t_{\mu\nu}(q)\, \frac{1}{q^2-M^2}\, -\, 
\ell_{\mu\nu}(q)\, \frac{\xi_Q}{q^2-\xi_Q M^2}\ ,
\end{equation}
with
\begin{displaymath} 
t_{\mu\nu}(q)\ =\ {}- g_{\mu\nu} + \frac{q_\mu q_\nu}{q^2}\, ,\quad
\ell_{\mu\nu}(q)\ =\ \frac{q_\mu q_\nu}{q^2}\ .
\end{displaymath}
One can write two separate DRs for the transverse self-energy, $\Pi_T$, of a
massive gauge boson, which crucially depend on the pole structure of Eq.\
(\ref{Prop1}), namely 
\begin{eqnarray}
\label{DRphys}
\Re e\bar{\Pi}_T(s) & = & 
\Re e\bar{\Pi}_T(M^2) + (s-M^2)\Re e\bar{\Pi}'_T(M^2)
+\frac{(s-M^2)^2}{\pi}\nonumber\\
&&\times \mbox{P}\, \int\limits_{\{ M^2_{phys}\} }^{+\infty} ds'
\, \frac{\Im m \bar{\Pi}_T (s')}{(s'-M^2)^2(s'-s)}\, ,\\
\label{DRunph}
\Re e\bar{\Pi}^{(\xi_Q)}_T (s) & = & (s-M^2)\Re e\bar{\Pi}'^{(\xi_Q)}_T(M^2)
+\frac{(s-M^2)^2}{\pi}\mbox{P}\, 
                              \int\limits_{\{ M^2_{unphys}\} }^{+\infty} ds'
\, \frac{\Im m \bar{\Pi}_T^{(\xi_Q)} (s')}{(s'-M^2)^2(s'-s)}\, .\nonumber\\
&&
\end{eqnarray}
In the first DR given in Eq.\ (\ref{DRphys}), the real part of $\Pi_T$, $\Re
e\bar{\Pi}_T$, is determined from branch cuts induced by physical poles, where
the masses of the real on-shell particles in the loop are collectively denoted
by $\{ M^2_{phys} \}$. In what follows we refer to such a DR as {\em physical}
DR. Note that $\Re e\bar{\Pi}_T$ depends only implicitly on the gauge choice.
In fact, $\Re e\bar{\Pi}_T$ can be viewed as the truncated part of the
self-energy that will survive if $\Re e{\Pi}_T$ is embedded in a $S$-matrix
element. In Eq.\ (\ref{DRunph}), the dispersive part of the two-point function
depends explicitly on $\xi_Q$-dependent unphysical thresholds, collectively
denoted by $\{ M^2_{unphys}\}$, which are induced by the longitudinal parts of
the gauge propagators contained in $\Im m\bar{\Pi}_T^{(\xi_Q)}$. Evidently,
one has the decomposition 
\begin{equation}
\Im m\Pi_T (s)\ =\ \Im m\bar{\Pi}_T(s) +\Im m\bar{\Pi}_T^{(\xi_Q)}(s)\, ,\quad
\Re e\Pi_T (s)\ =\ \Re e\bar{\Pi}_T(s) +\Re e\bar{\Pi}_T^{(\xi_Q)}(s)\, .
\end{equation}
${}$From Eq.\ (\ref{Prop1}), one can now isolate that part of the propagator 
that should be used in a physical DR. For $\xi_Q\not= 1$, one has 
\begin{equation} 
\Delta_{0\mu\nu}^{(\xi_Q)}\ \to \ U_{\mu\nu}(q)\ \equiv \
\Delta_{0\mu\nu }^{(\infty )}(q)\, .
\end{equation}
It is therefore obvious that the `physical' sector of an off-shell transition
amplitude in BFG (for $\xi_Q\not= 1$) ---or equivalently, the part of the
off-shell matrix element that satisfies a {\em physical} DR--- is effectively
obtained by considering all the internal propagators in the unitary gauge
($\xi_Q\to \infty $), but leaving the Feynman rules for the vertices in the
general $\xi_Q$ gauge. 

In view of a physical DR, the gauge $\xi_Q=1$ is very specific, since the
physical and unphysical poles coincide in such a case, making them
indistinguishable. At one-loop order, the results of this gauge are found to
collapse to those obtained via the PT~\cite{DDW}. Finally we remark in passing
that, if $\bar{\Pi}_T$ in $\xi_Q\not= 1$ is used for a definition of a
`physical' self-energy, one encounters problems with the high-energy unitarity
behaviour, even though the full $\Pi(\xi_{Q})$ is asymptotically well-behaved.
In the case of the one-loop $Z$ self-energy for example, for $\xi_{Q}\not= 1$
\cite{DDW}, $\bar{\Pi}_T$ contains terms proportional to $q^{4}$; all such
terms eventually cancel in the entire $\Pi(\xi_{Q})$ against the part that
contains the unphysical poles. Incidentally, it is interesting to notice that
the recovery of the correct asymptotic behaviour is the more delayed, {\em
i.e.}, it happens for larger values of $q^{2}$, the larger the value of
$\xi_{Q}$. However, if one was to resum only the $\bar{\Pi}_T$ part, the terms
proportional to $q^{4}$ would survive, leading to bad high energy behaviour.
If, on the other hand, one had resummed the full $\Pi(\xi_{Q})$, then one
would have introduced unphysical poles, as explained above. 

\setcounter{equation}{0}
\section {Unitarity and gauge invariance}
\indent

In this section, we will briefly discuss the basic field-theoretical
consequences resulting from the unitarity of the $S$-matrix theory, and
establish its connection with gauge invariance. In addition to the requirement
of explicit gauge invariance, the necessary conditions derived from unitarity
will constitute our guiding principle to analytically continue $n$-point
correlation functions in the off-shell region. Furthermore, we arrive at the
important conclusion that the resummed self-energies, in addition to being GFP
independent, must also be ``unitary'', in the sense that they do not spoil
unitarity when embedded in an $S$-matrix element.

The $T$-matrix element of a reaction $i\to f$ is defined via the relation
\begin{equation}
\langle f | S | i \rangle\ =\ \delta_{fi}\ +\ i(2\pi )^4
\delta^{(4)}(P_f - P_i)\langle f | T|i\rangle ,\label{tmatrix}
\end{equation}
where $P_i$ ($P_f$) is the sum of all initial (final) momenta of the
$|i\rangle$ ($| f \rangle$) state. Furthermore, imposing the unitarity
relation $S^\dagger S = 1$ leads to the OT: 
\begin{equation}
\langle f|T|i\rangle - \langle i |T|f\rangle^*\ =\
i\sum_{i'} (2\pi )^4\delta^{(4)}(P_{i'} - P_i)\langle i' | T | f \rangle^*
\langle i' | T | i \rangle. \label{optical}
\end{equation}
In Eq.~(\ref{optical}), the sum $\sum_{i'}$ should be understood to be over
the entire phase space and spins of all possible on-shell intermediate
particles $i'$. A corollary of this theorem is obtained if $i=f$. In this
particular case, we have 
\begin{equation}
\Im m \langle i|T|i\rangle\ =\ \frac{1}{2}
\sum_f (2\pi )^4\delta^{(4)}(P_f - P_i) |\langle f | T|i \rangle |^2.
\label{absorptive}
\end{equation}
In the conventional $S$-matrix theory with stable particles,
Eqs.~(\ref{optical}) and (\ref{absorptive}) hold also perturbatively. To be
precise, if one expands the transition $T=T^{(1)}+T^{(2)}+\cdots + T^{(n)} +
\cdots$, to a given order $n$, one has 
\begin{equation}
T^{(n)}_{fi}-T^{(n)*}_{if}\ =\ i\sum_{i'} (2\pi )^4
\delta^{(4)}(P_{i'}-P_i) \sum\limits_{k=1}^{n-1}
T^{(k)*}_{i'f} T^{(n-k)}_{i'i}.\label{optpert}
\end{equation}
There are two important conclusions that can be drawn from Eq.\
(\ref{optpert}). First, the anti-hermitian part of the LHS of Eq.\
(\ref{optpert}) contains, in general, would-be Goldstone bosons or ghost fields
\cite{FP}. Such contributions manifest themselves as Landau singularities at
unphysical points, {\em e.g.}, $q^2=\xi_Q M_W^2$ for a $W$ propagator in a
general BFG. However, unitarity requires that these unphysical contributions
should vanish, as can be read off from the RHS of Eq.\ (\ref{optpert}). Second,
the RHS explicitly shows the connection between gauge invariance and unitarity
at the quantum loop level. To lowest order for example, the RHS consists of the
product of GFP independent on-shell tree amplitudes, thus enforcing the
gauge-invariance of the imaginary part of the one-loop amplitude on the LHS.

The above powerful constraints imposed by unitarity will be in effect as long
as one computes {\em full} amplitudes to a finite order in perturbation
theory. However, for resummation purposes, a certain sub-amplitude, {\em
i.e.}, a part of the full amplitude, must be singled out and subsequently
undergo a Dyson summation, while the rest of the $S$-matrix is computed to a
finite order $n$. Therefore, if the resummed amplitude contains gauge
artifacts and/or unphysical thresholds, the cancellations imposed by Eq.\
(\ref{optpert}) will only operate up to order $n$, introducing unphysical
contributions of order $n+1$ or higher. To avoid the contamination of the
physical amplitudes by such unphysical artifacts, we impose the following two
requirements on the effective Green's functions, when one attempts to continue
them analytically in the off-shell region for the purpose of resummation: 

\begin{itemize}

\item[(i)] The off-shell $n$-point correlation functions ought to be derivable
from or embeddable into $S$-matrix elements. 

\item[ (ii)] The off-shell Green's functions should not display unphysical
thresholds induced by unphysical Landau singularities, as has been described
above. 

\end{itemize}

Even though property (i) is automatic for Green's functions generated by the
functional differentiation of the conventional path-integral functional, in
general the off-shell amplitudes so obtained fail to satisfy property (ii). In
the PT framework instead, both conditions are satisfied: effective Green's
functions are directly derived from the $S$-matrix amplitudes (so condition
(i) is satisfied by construction) and contain only physical thresholds, so
that unitarity is not explicitly violated \cite{JP&AP}. 

In our discussion of unitarity at one-loop, we will make 
extensive use of the following
two-body Lorentz-invariant phase-space (LIPS) integrals: 
The scalar integral
\begin{eqnarray}
\int dX_{LIPS} &=& \frac{1}{(2\pi)^2}\, 
\int d^{4}k_1 \int d^{4}k_2\, \delta_+ (k^2_1-m^2_1)\delta_+ (k^2_2-m^2_2)
\delta^{(4)}(q-k_1-k_2)
\nonumber\\
&=& \theta(q^0)\theta [q^2-{(m_1+m_2)}^2]\, \frac{1}{8\pi\, q^2}\,
\lambda^{1/2} (q^2,m_1^2,m_2^2)\, ,
\label{LIPS1}
\end{eqnarray}
where $\lambda (x,y,z)= (x-y-z)^2-4yz$ and $\delta_+(k^2-m^2)\equiv \theta
(k^0)\delta(k^2-m^2)$, and the  
tensor integral: 
\begin{eqnarray}
\int dX_{LIPS} {(k_1-k_2)}_{\mu}{(k_1-k_2)}_{\nu} &=&
\Big\{ 
\frac{\lambda (q^2,m_1^2,m_2^2)}{3q^2}\, t_{\mu\nu}(q) +\, 
\Big[ \frac{\lambda (q^2,m_1^2,m_2^2)}{q^2} - q^2\nonumber\\
&&+ 2 (m^2_1+m^2_2) \Big]\, \ell_{\mu\nu}(q)\, \Big\} \times \int dX_{LIPS}\, .
\label{LIPS2}
\end{eqnarray}
The Lorentz projection tensors, $t_{\mu\nu}(q)$ and $\ell_{\mu\nu}(q)$,
have been defined after Eq.\ (\ref{Prop1}).

\setcounter{equation}{0}
\section {The case of QCD}
\indent

In this section, we show that a self-consistent picture may be obtained by
resorting to such fundamental properties of the $S$-matrix as unitarity and
analyticity, using as additional input only EWIs for tree-level, on-shell
processes, and tree-level vertices and propagators.  It is important to
emphasize that the GFP independence of the results emerges {\em automatically}
from the previous considerations.

We begin from the RHS of the optical relation given in Eq.\ 
(\ref{absorptive}). The RHS involves on-shell physical processes, which
satisfy the EWIs. It turns out that the full exploitation of those EWIs leads
unambiguously to a decomposition of the tree-level amplitude into
propagator-, vertex- and box-like structures. The propagator-like structure
corresponds to the imaginary part of the effective propagator under
construction. By imposing the additional requirement that the effective
propagator be an analytic function of $q^2$ one arrives at a DR, which,
up to renormalization-scheme choices, leads to a unique result for the real 
part. 

Consider the forward scattering process $q {\bar q}\rightarrow q{\bar q}$. From
the OT, we then have
\begin{equation}
\Im m \langle q\bar{q}|T|q\bar{q}\rangle\ =\ \frac{1}{2}\, 
\left( \frac{1}{2} \right)\, 
\int dX_{LIPS}\,  
\langle q\bar{q}|T|gg\rangle \langle gg|T|q\bar{q}\rangle^{*}\, .
\label{OTgg}
\end{equation}
In Eq.\ (\ref{OTgg}), the statistical factor 1/2 in parentheses arises from
the fact that the final on-shell gluons should be considered as identical
particles in the total rate. We now set ${\cal M}=\langle
q\bar{q}|T|q\bar{q}\rangle $ and ${\cal T}=\langle q\bar{q}|T|gg\rangle$, and
focus on the RHS of Eq.\ (\ref{OTgg}). Diagrammatically, the amplitude ${\cal
T}$ consists of two distinct parts: $t$ and $u$-channel graphs that contain an
internal quark propagator, ${{\cal T}_{t}}^{ab}_{\mu\nu}$, as shown in Figs.\
3(a) and 3(b), and an $s$-channel amplitude, ${{\cal T}_{s}}^{ab}_{\mu\nu}$,
which is given in Fig.\ 3(c). The subscript ``$s$'' and ``$t$'' refers to the
corresponding Mandelstam variables, {\em i.e.}\ $s=q^2=
(p_1+p_2)^2=(k_1+k_2)^2$, and $t=(p_1-k_1)^2=(p_2-k_2)^2$. 
Defining 
\begin{equation}
V_{\rho}^{c}\ =\ g\bar{v}(p_2)\, \frac{\lambda^c}{2}\gamma_{\rho}\, u(p_1)\, ,
\end{equation}
we have that
\begin{equation}
{\cal T}^{ab}_{\mu\nu}={{\cal T}_{s}}^{ab}_{\mu\nu} (\xi )+
{{\cal T}_{t}}^{ab}_{\mu\nu}\, ,
\label{DefT}
\end{equation}
with
\begin{eqnarray}
{{\cal T}_{s}}^{ab}_{\mu\nu}(\xi )& =&
-gf^{abc}\, \Delta^{(\xi ),\rho\lambda}_0(q)
\Gamma_{\lambda\mu\nu}(q,-k_1,-k_2)\, V_{\rho}^{c}\, ,
\label{Ts}\\
{{\cal T}_{t}}^{ab}_{\mu\nu} &=& -ig^2\bar{v}(p_2)\Big( 
\, \frac{\lambda^b}{2}\gamma^{\nu}\, \frac{1}{\not\! p_1-\not\! k_1 - m}
\, \frac{\lambda^a}{2}\gamma^{\mu}\ +\
\frac{\lambda^a}{2}\gamma^{\mu}\,
\frac{1}{\not\! p_1-\not\! k_2-m}\, 
\gamma^{\nu}\frac{\lambda^b}{2}\, \Big)u(p_1)\, ,\qquad 
\label{Tt}
\end{eqnarray}
where
\begin{equation}
\Gamma_{\lambda\mu\nu}(q,-k_1,-k_2)\ =\ 
(k_1-k_2)_{\lambda}g_{\mu\nu}\, +\, (q+k_2)_{\mu}g_{\lambda\nu}\,
-\, (q+k_1)_{\nu}g_{\lambda\nu}\, .
\label{3GV}
\end{equation}
\begin{center}
\begin{picture}(340,300)(0,0)
\SetWidth{0.8}

\ArrowLine(20,250)(50,250)\Text(25,260)[r]{$q (p_1)$}
\ArrowLine(50,250)(50,210)\Text(55,230)[l]{$q$}
\ArrowLine(50,210)(20,210)\Text(25,200)[r]{$\bar{q} (p_2)$}
\Gluon(50,250)(80,250){3}{3}\Text(120,265)[r]{$g (k_1),\mu , a$}
\Gluon(50,210)(80,210){3}{3}\Text(120,200)[r]{$g (k_2),\nu , b$}
\Text(50,160)[l]{{\bf (a)}}

\ArrowLine(140,250)(170,250)
\ArrowLine(170,250)(170,210)\Text(165,230)[r]{$q$}
\ArrowLine(170,210)(140,210)
\Gluon(170,250)(200,210){3}{5}
\Gluon(170,210)(200,250){3}{5}
\Text(170,160)[l]{{\bf (b)}}

\ArrowLine(240,250)(270,230)
\ArrowLine(270,230)(240,210)
\Gluon(270,230)(300,230){3}{3}\Text(270,240)[l]{$\lambda , c$}
\Gluon(300,230)(330,250){3}{3}
\Gluon(300,230)(330,210){3}{3}
\Text(280,160)[l]{{\bf (c)}}

\ArrowLine(120,100)(150,80)
\ArrowLine(150,80)(120,60)
\Gluon(150,80)(180,80){3}{3}\Text(150,90)[l]{$\lambda ,c$}
\DashArrowLine(180,80)(210,100){1}\Text(215,100)[l]{$c_g ,a$}
\DashArrowLine(210,60)(180,80){1}\Text(215,60)[l]{$\bar{c}_g ,b$}
\Text(160,30)[l]{{\bf (d)}}

\end{picture}\\
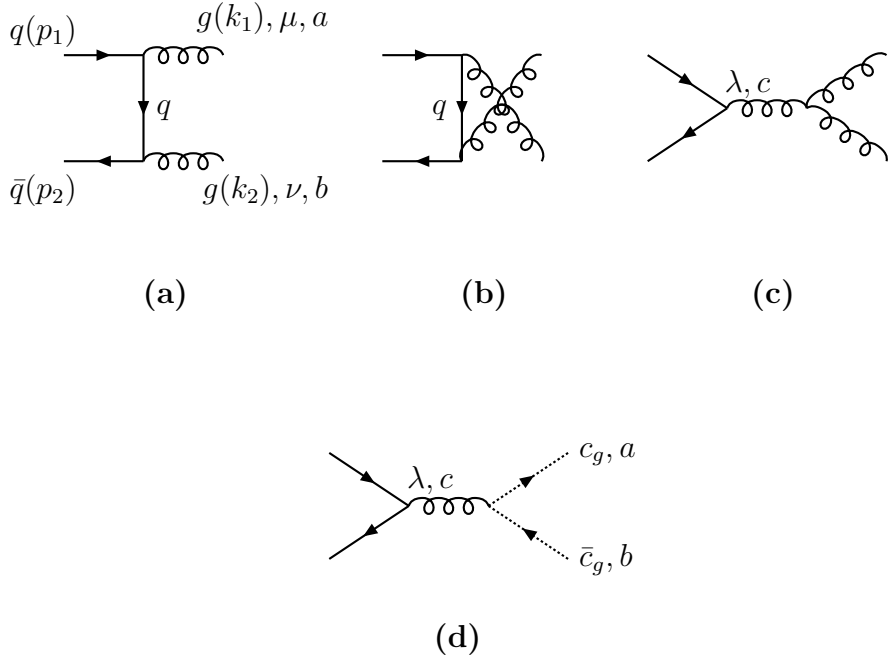
{\bf Fig.\ 3:} Diagrams (a)--(c) contribute to ${\cal T}^{ab}_{\mu\nu}$,
and diagram (d) to ${\cal S}^{ab}$.
\end{center}
Notice that ${\cal T}_{s}$ depends explicitly on the GFP $\xi$, through the
tree-level gluon propagator $\Delta^{(\xi)}_{0\mu\nu}(q)$, whereas ${\cal
T}_{t}$ does not. The explicit expression of $\Delta^{(\xi )}_{0\mu\nu}(q)$
depends on the specific gauge fixing procedure chosen.  In addition, we define
the quantities ${\cal S}^{ab}$ and ${\cal R}^{ab}_{\mu}$ as follows:
\begin{eqnarray}
{\cal S}^{ab}&=&gf^{abc}\, \frac{k^\sigma_1}{q^2}\, 
V_{\sigma}^{c}\nonumber\\
&=&-gf^{abc}\, \frac{k^\sigma_2}{q^2}\, V_{\sigma}^{c}
\label{S}
\end{eqnarray}
and 
\begin{equation}
{\cal R}_{\mu}^{ab}\ =\ gf^{abc}\, V_{\mu}^{c}\, .
\label{R}
\end{equation}
Clearly, 
\begin{equation}
k_1^{\sigma}{\cal R}_{\sigma}^{ab}\ =\
-k_2^{\sigma}{\cal R}_{\sigma}^{ab}\ =\ q^2{\cal S}^{ab}.
\label{D2}
\end{equation}
We then have
\begin{eqnarray}
\Im m {\cal M}&=& \frac{1}{4}\, {\cal T}^{ab}_{\mu\nu}\, P^{\mu\sigma}
(k_1,\eta_1)\, P^{\nu\lambda}(k_2,\eta_2)\, {\cal T}^{ab*}_{\sigma\lambda}
\nonumber\\
&=& \frac{1}{4}\Big[ {{\cal T}_{s}}^{ab}_{\mu\nu}(\xi)+
{{\cal T}_{t}}^{ab}_{\mu\nu}\Big]\, P^{\mu\sigma}(k_1,\eta_1)\,
P^{\nu\lambda}(k_2,\eta_2)\, \Big[ {{\cal T}_{s}}^{ab*}_{\sigma\lambda}(\xi)
+{{\cal T}_{t}}^{ab*}_{\sigma\lambda}\Big],
\label{MM}
\end{eqnarray}
where the polarization tensor $P^{\mu\nu}(k,\eta )$ is given by
\begin{equation}
P_{\mu\nu}(k,\eta )\ =\ -g_{\mu\nu}+ \frac{\eta_{\mu}k_{\nu}
+\eta_{\nu}k_{\mu} }{\eta k} + 
\eta^2 \frac{k_{\mu}k_{\nu}}{{(\eta k)}^2}\, .
\label{PhotPol}
\end{equation}
Moreover, we have that on-shell, {\em i.e.}, for $k^{2}=0$,
$k^{\mu}P_{\mu\nu}=0$. By virtue of  this last property, we see immediately
that if we write the three-gluon vertex of Eq.\ (\ref{3GV}) in the form 
\begin{eqnarray}
\Gamma_{\lambda\mu\nu} (q,-k_1,-k_2) &=& [(k_1-k_2)_{\lambda}g_{\mu\nu}+
2q_{\mu}g_{\lambda\nu}-2q_{\nu}g_{\lambda\mu}]\
+\ (-k_{1\mu}g_{\lambda\nu}+k_{2\nu}g_{\lambda\mu}) \nonumber\\
&=& \Gamma^F_{\lambda\mu\nu}(q,-k_1,-k_2)\ +\ 
\Gamma^P_{\lambda\mu\nu}(q,-k_1,-k_2)\, ,
\label{GFGP}
\end{eqnarray}
the term $\Gamma^P_{\rho\mu\nu}$ dies after hitting the polarization vectors
$P_{\mu\sigma}(k_1,\eta_1)$ and $P_{\nu\lambda}(k_2,\eta_2)$.  Therefore, if
we denote by ${\cal T}_{s}^{F}(\xi)$ the part of ${\cal T}_{s}$ which
survives, Eq.\ (\ref{MM}) becomes 
\begin{equation}
\Im m {\cal M}\ =\ \frac{1}{4}\,
\big[ {\cal T}_{s}^{F}(\xi)+{\cal T}_{t}\big]^{ab}_{\mu\nu}\,
P^{\mu\sigma}(k_1,\eta_1 )\, P^{\nu\lambda}(k_2,\eta_2  )\,
\big[ {\cal T}_{s}^{F}(\xi)
+{\cal T}_{t} \big]^{ab*}_{\sigma\lambda}\, .
\label{MM22}
\end{equation}
The next step is to verify that any dependence on the GFP inside the
propagator $\Delta^{(\xi)}_{0\mu\nu}(q)$ of the off-shell gluon will
disappear. This is indeed so, because the longitudinal parts of
$\Delta_{0\mu\nu}$ either vanish because the external quark current is
conserved, or because they trigger the following EWI: 
\begin{equation}
q^{\mu}\Gamma^{F}_{\mu\alpha\beta}(q, -k_1, -k_2)\ =\ 
(k_1^2\ -\ k_2^2)g_{\alpha\beta}\, ,
\label{FWI}
\end{equation}
which vanishes on shell.  This last EWI is crucial, because in general,
current conservation alone is not sufficient to guarantee the GFP independence
of the final answer. In the covariant gauges for example, the gauge fixing term
is proportional to $q^{\mu}q^{\nu}$; current conservation kills such a term. 
But if we had chosen an axial gauge instead, {\em i.e.} 
\begin{equation}
\label{Deleta}
\Delta^{(\tilde{\eta})}_{0\mu\nu}(q)\ =\ \frac{P_{\mu\nu}
(q,\tilde{\eta})}{q^2}\, , 
\end{equation}
where $\tilde{\eta} \neq \eta$ in general, then only the term
${\tilde{\eta}_{\nu}}q_{\mu}$ vanishes because of current conservation,
whereas the term ${\tilde{\eta}_{\nu}}q_{\mu}$ can only disappear if Eq.\
(\ref{FWI}) holds.  So, Eq.\ (\ref{MM22}) becomes
\begin{equation}
\Im m {\cal M}\ =\ \frac{1}{4}
({\cal T}_{s}^{F}+{\cal T}_{t})^{ab}_{\mu\nu}\,
P^{\mu\sigma}(k_1,\eta_1)\, P^{\nu\lambda}(k_2,\eta_2)\,
({\cal T}_{s}^{F}+{\cal T}_{t})^{ab*}_{\sigma\lambda}\, ,
\label{MM2}
\end{equation}
where the GFP-{\em independent} quantity ${\cal T}_{s}^{F}$ is given by
\begin{equation}
{{\cal T}_{s}}^{F,ab}_{\mu\nu}\ =\
-gf^{abc}\, \frac{g^{\rho\lambda}}{q^2}\, \,
\Gamma^{F}_{\lambda\mu\nu}(q,-k_1,-k_2)\, V_{\rho}^{c}\, .
\label{TsF}
\end{equation}
Next, we want to show that the dependence on $\eta_{\mu}$ and $\eta^2$
stemming from the polarization vectors disappears.  Using the on shell
conditions $k_1^2=k_2^2=0$, we can easily verify the following EWIs: 
\begin{eqnarray}
k_1^{\mu}{{\cal T}_{s}}^{F,ab}_{\mu\nu} & = &  
2 k_{2\nu}{\cal S}^{ab}\, -\, {\cal R}_{\nu}^{ab} \, ,
\label{w1}\\
k_2^\nu {{\cal T}_{s}}^{F,ab}_{\mu\nu}  & = &  2 k_{1\mu}{\cal S}^{ab} 
\, +\, {\cal R}_{\mu}^{ab}  \, ,\label{w2}\\
k_1^{\mu}{{\cal T}_{t}}^{ab}_{\mu\nu} & = & {\cal R}_{\nu}^{ab} \, ,
\label{w3}\\
k_2^{\nu}{{\cal T}_{t}}^{ab}_{\mu\nu} & = & -{\cal R}_{\mu}^{ab} 
\label{w4}\, ,
\end{eqnarray}
from which we have that 
\begin{eqnarray}
k_1^{\mu}k_2^{\nu}{{\cal T}_{s}}^{F,ab}_{\mu\nu}& = & q^2{\cal S}^{ab}
\label{w5}\, , \\
k_1^{\mu}k_2^{\nu}{{\cal T}_{t}}^{ab}_{\mu\nu} &=& -q^2{\cal S}^{ab}\, .
\label{w6}
\end{eqnarray}
Using the above EWIs, it is now easy to check that indeed, all dependence on
both $\eta_{\mu}$ and $\eta^2$ cancels in Eq.\ (\ref{MM2}), as it should, and
we are finally left with (omitting the fully contracted colour and Lorentz
indices): 
\begin{eqnarray}
\Im m {\cal M} &=& \frac{1}{4}\, \Big[
\Big( {\cal T}_{s}^{F}{{\cal T}_{s}^{F}}^{*} -8 {\cal S}{\cal S}^{*}\Big) 
+ \Big( {\cal T}_{s}^{F}{\cal T}_{t}^{*} + {{\cal T}_{s}^{F}}^{*}
{\cal T}_{t} \Big) + {\cal T}_{t}{\cal T}_{t}^{*} \Big]\nonumber\\
&=& \Im m \widehat{{\cal M}}_1+ \Im m \widehat{{\cal M}}_2+
\Im m \widehat{{\cal M}}_{3} \, .
\label{MM3}
\end{eqnarray}
The first part is the genuine propagator-like piece, the second is the vertex,
and the third the box.  Employing the fact that
\begin{equation}
\Gamma^{F}_{\rho\mu\nu}\Gamma^{F,\mu\nu}_{\lambda} \ =\ 
-8q^2t_{\rho\lambda}(q) +
4{(k_1-k_2)}_{\rho}{(k_1-k_2)}_{\lambda}
\label{FRE}
\end{equation}
and 
\begin{eqnarray}
{\cal S}{\cal S }^{*} &=& g^2\, c_A\, V^c_\rho \, 
\frac{k_1^{\rho}k_1^{\lambda}}{(q^2)^2}\, V^{c}_{\lambda} \nonumber\\
&=& \frac{g^2}{4}\, c_A\, V^c_{\rho}\, \frac{(k_1-k_2)^\rho 
(k_1-k_2)^\lambda }{(q^2)^2}\, V^{c}_{\lambda}\, ,
\end{eqnarray}
where $c_{A}$ is the eigenvalue of the Casimir operator in the adjoint
representation ($ c_A=N$ for SU$(N)$), we obtain for $\Im m \widehat{{\cal
M}}_1$
\begin{equation}
\Im m \widehat{{\cal M}}_1\ =\ \frac{g^2}{2}\, c_A
V^{c}_{\mu}\, \frac{1}{q^2}\, \Big[ -4q^2t^{\mu\nu}(q)\, +\,
{(k_1-k_2)}^{\mu}{(k_1-k_2)}^{\nu}\Big]\, \frac{1}{q^2}\, V^{c}_{\nu}\, .
\end{equation} 
This last expression must be integrated over the available phase space. With
the help of Eqs.\ (\ref{LIPS1}) and (\ref{LIPS2}), we arrive at the final
expression
\begin{equation}
\Im m \widehat{{\cal M}}_1\ =\
V^c_\mu\, \frac{1}{q^2}\, \Im m \widehat{\Pi}^{\mu\nu}(q)\,
\frac{1}{q^2}\, V^c_\nu\, ,
\end{equation}
with  
\begin{equation}
\label{IMQCD}
\Im m \widehat{\Pi}_{\mu\nu}(q)\ =\ -\, \frac{\alpha_s}{4}\, 
\frac{11c_A}{3}\, q^2 t_{\mu\nu}(q)\, ,
\end{equation}
and $\alpha_s=g^2/(4\pi)$. 

Before we proceed, we make the following remark. It is well-known that the
vanishing of the longitudinal part of the gluon self-energy is an important
consequence of gauge invariance. One might naively expect that even if a
non-vanishing longitudinal part had been induced by some contributions which
do not respect gauge invariance, it would not have contributed to physical
processes, since the gluon self-energy couples to conserved fermionic
currents, thus projecting out only the transverse degrees of the gluon vacuum
polarization. However, this expectation is not true in general. Indeed, if one
uses, for example, the tree-level gluon propagator in the axial gauge, as given
in Eq.\ (\ref{Deleta}), then there will be residual $\eta$-dependent terms
induced by the longitudinal component of the gluon vacuum polarization, which
would not vanish, despite the fact that the external quark currents are
conserved. Such terms are obviously gauge dependent. Evidently, projecting out
only the transverse parts of Green's functions will not necessarily render
them gauge invariant. 

The vacuum polarization of the gluon within the PT is given by \cite{PT}
\begin{equation}
\label{PTgg}
\widehat{\Pi}_{\mu\nu}(q)\ =\ \frac{\alpha_s}{4\pi}\, \frac{11c_A}{3}\, 
t_{\mu\nu}(q)\, q^2\, \Big[\, \ln\Big(-\frac{q^2}{\mu^2}\Big)\, +\, C_{UV}\, 
\Big]\, .
\end{equation}
Here, $C_{UV}=1/\epsilon -\gamma_E + \ln 4\pi + C$, with $C$ being some
constant and $\mu$ is a subtraction point. In Eq.\ (\ref{PTgg}), it is
interesting to notice that a change of $\mu^2\to \mu '^2$ gives rise to a
variation of the constant $C$ by an amount $C'-C=\ln \mu'^2/\mu^2$.  Thus, a
general $\mu$-scheme renormalization yields
\begin{eqnarray}
\label{RPTgg}
\widehat{\Pi}_T^R (s) &=& \widehat{\Pi}_T (s)\, -\, (s-\mu^2)\Re e
\widehat{\Pi}_T'(\mu^2)\, -\, \Re e\widehat{\Pi}_T(\mu^2 ) \nonumber\\
&=& \frac{\alpha_s}{4\pi}\, \frac{11c_A}{3}\, s\, \Big[
\ln\Big(-\frac{s}{\mu^2}\Big)\, -\, 1\, +\, \frac{\mu^2}{s}\, \Big]\, .
\end{eqnarray}
${}$From Eq.\ (\ref{DR3}), one can readily see that $\Re e
\widehat{\Pi}^R_T(s)$ can be calculated by the following double 
subtracted DR:
\begin{equation}
\label{QCDDR}
\Re e \widehat{\Pi}^R_T(s)\ =\ \frac{(s-\mu^2)^2}{\pi}\ \mbox{P}\,
\int\limits_0^\infty\, ds' \frac{\Im m \widehat{\Pi}_T(s')}{(s'-\mu^2)^2 
(s'-s)}\, .
\end{equation}
Inserting Eq.\ (\ref{IMQCD}) into Eq.\ (\ref{QCDDR}), it is not difficult to
show that it leads to the result given in Eq.\ (\ref{RPTgg}), a fact that 
demonstrates the analytic power of the DR.

It is important to emphasize that the above derivation rigorously proves the
GFP independence of the one-loop PT effective Green's functions, for {\em
every} gauge fixing procedure. Indeed, in our derivation, we have solely
relied on the RHS of the OT, which we have rearranged in a well-defined way,
{\em after} having explicitly demonstrated its GFP-independence. The proof of
the GFP-independence of the RHS presented here is, of course, expected on
physical grounds, since it only relies on the use of EWIs, triggered by the
longitudinal parts of the gluon tree-level propagators. Note that the
tree-level tri-gluon coupling, $\Gamma_{\lambda\mu\nu}$, is uniquely given by
Eq.\ (\ref{3GV}).  Since the GFP-dependence is carried entirely by the
longitudinal parts of the gluon tree-level propagator in {\em any}
gauge-fixing scheme whereas the $g^{\mu\nu}$ part is GFP-independent and 
universal, the proof presented here is generally true. Obviously, the final
step of reconstructing the real part from the imaginary by means of a DR does
not introduce any gauge-dependences. 
  
\setcounter{equation}{0}
\section{The QCD analysis from BRS considerations}
\indent

In this section, we will show how we can obtain the same answer by resorting
only to the EWIs that one obtains as a direct consequence of the BRS symmetry
of the quantum Lagrangian. 

If we consider ${\cal T}_{\mu\nu}^{ab}$ as before, it is easy to show that 
it satisfies the following BRS identities \cite{ChengLi}:
\begin{eqnarray}
k^{\mu}_1{\cal T}_{\mu\nu}^{ab} &=& k_{2\nu}{\cal S}^{ab}\, ,\nonumber\\
k^{\nu}_2{\cal T}_{\mu\nu}^{ab} &=& k_{1\mu}{\cal S}^{ab}\, ,  \nonumber\\
k^{\mu}_1k^{\nu}_2{\cal T}_{\mu\nu}^{ab} &=& 0\, ,
\label{BRSg}
\end{eqnarray}
where $S^{ab}$ is the ghost amplitude shown in Fig.\ 3(d); its closed form is
given in Eq.\ (\ref{S}).

Notice that the BRS identities of Eq.\ (\ref{BRSg}) are different from those
listed in Eqs. (\ref{w1})--(\ref{w6}), because  the term
$\Gamma_{\mu\nu\rho}^{P}$ had been removed in the latter case. Here, we follow
a different sequence and do not kill the term $\Gamma_{\mu\nu\rho}^{P}$;
instead, we will exploit the {\em exact} BRS identities from the very
beginning.

We start again with the expression for $\Im m {\cal M}$ given in Eq.\
(\ref{MM}). First of all, it is easy to verify again that the dependence on
the GFP of the off-shell gluon vanishes. This is so because of the tree-level
EWI, involving the {\em full} vertex $\Gamma_{\mu\nu\rho}$,
\begin{equation}
q^{\lambda}\Gamma_{\lambda\mu\nu}(q,-k_1,-k_2)\ =\ 
k^2_2\, t_{\mu\nu}(k_2 )\ -\ k^2_1\, t_{\mu\nu}(k_1 )\, .
\end{equation}
The RHS vanishes after contracting with the polarization vectors, and
employing the on-shell condition $k^2_1=k^2_2=0$.  Again, by virtue of the BRS
identities and the on-shell condition $k_1^2=k_2^2=0$, the dependence of $\Im m
{\cal M}$ on the parameters $\eta_{\mu}$ and $\eta^2$ cancels, and we eventually
obtain
\begin{eqnarray}
\Im m {\cal M} &=& \frac{1}{4}\,  {\cal T}_{\mu\nu}\, 
P^{\mu\rho}(k_1,\eta_1)\, P^{\nu\sigma}(k_2,\eta_2)\, 
{\cal T}_{\rho\sigma}^* 
\nonumber\\
&=& \frac{1}{4}\, \Big( {\cal T}^{\mu\nu} {\cal T}_{\mu\nu}^{*}\ -\ 
2 {\cal S}{\cal S}^{*}\Big) \nonumber\\
&=& \frac{1}{4}\, 
\Big[({\cal T}_{s}^{F}+{\cal T}_{s}^{P}+{\cal T}_{t})^{\mu\nu}
({\cal T}_{s}^{F}+{\cal T}_{s}^{P}+{\cal T}_{t})_{\mu\nu}^{*}\ -\ 2 
{\cal S}{\cal S}^{*}\Big]\, ,
\label{GIFA}
\end{eqnarray}
where
\begin{equation}
{{\cal T}_{s}}^{P,ab}_{\mu\nu}=
-gf^{abc}\, \frac{g^{\rho \lambda}}{q^2}\,
\Gamma^{P}_{\lambda\mu\nu}(q,-k_1,-k_2)\, V_{\rho }^{c}\, .
\label{TsP}
\end{equation}

At this point, one  must recognize that due to the four-momenta of the
trilinear vertex $\Gamma^{P}$ inside ${\cal T}_{s}^{P}$, one can further
trigger the EWIs, exactly as one did in order to derive from Eq.\ (\ref{MM})
the last step of Eq.\ (\ref{GIFA}). In fact, only the process-independent
terms contained in $\Im m {\cal M}$ will be projected out on account of the
BRS identities of Eq.\ (\ref{BRSg}).  It is important to emphasize that ${\cal
T}_{s}^{F}$ and ${\cal T}_{t}$ do not contain any pinching momenta. This is
particular to this example, where we have only two gluons as final states, but
is not true for more gluons.  To further exploit the EWIs derived from BRS
symmetries, we re-write the RHS of Eq.\ (\ref{GIFA}) in the following way (we
omit the fully contracted Lorentz indices): 
\begin{eqnarray}
\Im m {\cal M} &=& \frac{1}{4}\, 
\Big[({\cal T}_{t}+{\cal T}^{P}_{s}+{\cal T}^{F}_{s})
({\cal T}_{t}+{\cal T}^{P}_{s}+{\cal T}^{F}_{s})^{*}\ -\ 2 
{\cal S}{\cal S}^{*} \Big]
\nonumber\\
&=& \frac{1}{4}\, \Big[({\cal T}^{F}_{s}{{\cal T}^{F}_{s}}^{*}
-{\cal T}^{P}_{s}{{\cal T}^{P}_{s}}^{*}
+{\cal T}^{P}_{s}{\cal T}^{*}+{\cal T}{{\cal T}^{P}_{s}}^{*} - 2 {\cal S}
{\cal S}^{*})
+( {\cal T}_{t} {{\cal T}^{F}_{s}}^{*}
+{\cal T}^{F}_{s}{\cal T}_{t}^{*} )
+{\cal T}_{t}{\cal T}_{t}^{*}\Big]
\nonumber\\
&=& \Im m \widehat{{\cal M}}_1+\Im m \widehat{{\cal M}}_2+
\Im m \widehat{{\cal M}}_{3}\, .
\label{makrynari}
\end{eqnarray}
In Eq.\ (\ref{makrynari}), the reader may recognize the rearrangement
characteristic of the ``intrinsic'' PT, presented in \cite{CornPap}.

Inserting the explicit form of ${\cal T}^{P}_s$ given in Eq.\ (\ref{TsP}) into
Eq.\ (\ref{makrynari}) and using the BRS identities, 
\begin{eqnarray}
\label{TPsid}
{\cal T}^{P}_{s} {\cal T}^{*} &=& -2 {\cal S}{\cal S}^{*}\, ,\nonumber\\
{\cal T}^{P}_{s}{\cal T}^{P*}_{s} &=& 2 {\cal S}{\cal S}^{*}\, ,
\end{eqnarray}
we obtain
\begin{eqnarray}
\Im m \widehat{{\cal M}}_1 &= & \frac{1}{4}\, \Big( 
{\cal T}^{F}_{s}{{\cal T}^{F}_{s}}^{*}-{\cal T}^{P}_{s}{{\cal T}^{P}_{s}}^{*}
+{\cal T}^{P}_{s}{\cal T}^{*}+{{\cal T}^{P}_{s}}^{*}{\cal T}
-2{\cal S}{\cal S}^{*}\Big)\nonumber\\
&=& \frac{1}{4}\, \Big(
{\cal T}^{F}_{s}{{\cal T}^{F}_{s}}^{*}-8{\cal S}{\cal S}^{*}\Big)\, ,
\label{Final}
\end{eqnarray}
which is the same result found in the previous section, {\em i.e.},  Eq.\
(\ref{MM3}). 

An interesting by-product of the above analysis is that one is able to show
the independence of the PT results  of the number of the external
fermionic currents \cite{NJW}. Indeed, the BRS identities in Eqs.\
(\ref{BRSg}), as well as those given in Eq.\ (\ref{TPsid}), will still hold
for any transition amplitude of $n$-fermionic currents to two gluons.  By
analogy, one can decompose the transition amplitude into ${\cal T}_t$ and
${\cal T}_s$ structures. Similarly, the form of the sub-structures ${\cal
T}^F_s$ and ${\cal T}^P_s$ will then change accordingly. In fact, the only
modification will be that the vector current, $V^c_\rho$, contained in Eqs.\
(\ref{TsF}) and (\ref{TsP}) will now represent the transition of one gluon to
$n$-fermionic currents. Making use of the ``intrinsic'' PT, one then obtains
the result given in Eq.\ (\ref{Final}). Hence, we can conclude that the PT
does not depend on the number of the external fermionic currents attached to
gluons.

\setcounter{equation}{0}
\section{The electroweak case}
\indent

In this section, we will show how the same considerations apply directly to the
case of the electroweak sector of the SM. We consider the charged current
process $e^- \nu \to e^- \nu$ and assume that the electron mass $m_e$ is
non-zero, so that the external current is not conserved.  We focus on the part
of the amplitude which has a threshold at $q^2=M^2_{W}$. This corresponds the
virtual process $W^-\to W^-\gamma$, where $\gamma$ is the photon. From the
OT, we have 
\begin{equation}
\Im m \langle e^{-}\nu|T|e^{-}\nu\rangle\ =\ \frac{1}{2}
\int dX_{LIPS}\ \langle e^{-}\nu|T|W^{-}\gamma\rangle 
\langle W^{-}\gamma |T|e^{-}\nu\rangle^{*}\, .
\label{OTWg}
\end{equation}
\begin{center}
\begin{picture}(340,200)(0,0)
\SetWidth{0.8}

\ArrowLine(20,150)(50,150)\Text(25,160)[r]{$e (p_1)$}
\ArrowLine(50,150)(50,110)\Text(55,130)[l]{$e$}
\ArrowLine(50,110)(20,110)\Text(25,100)[r]{$\bar{\nu} (p_2)$}
\Photon(50,150)(80,150){2}{3}\Text(110,160)[r]{$\gamma (k_2),\nu $}
\Photon(50,110)(80,110){2}{3}\Text(110,100)[r]{$W^- (k_1),\mu $}
\Text(50,50)[l]{{\bf (a)}}

\ArrowLine(120,150)(150,130)
\ArrowLine(150,130)(120,110)
\DashArrowLine(150,130)(180,130){2}\Text(155,140)[l]{$G^-$}
\Photon(180,130)(210,150){2}{3}
\Photon(180,130)(210,110){2}{3}
\Text(155,50)[l]{{\bf (b)}}

\ArrowLine(250,150)(280,130)
\ArrowLine(280,130)(250,110)
\Photon(280,130)(310,130){2}{3}\Text(280,140)[l]{$W^-,\rho$}
\Photon(310,130)(340,150){2}{3}
\Photon(310,130)(340,110){2}{3}
\Text(285,50)[l]{{\bf (c)}}

\end{picture}\\
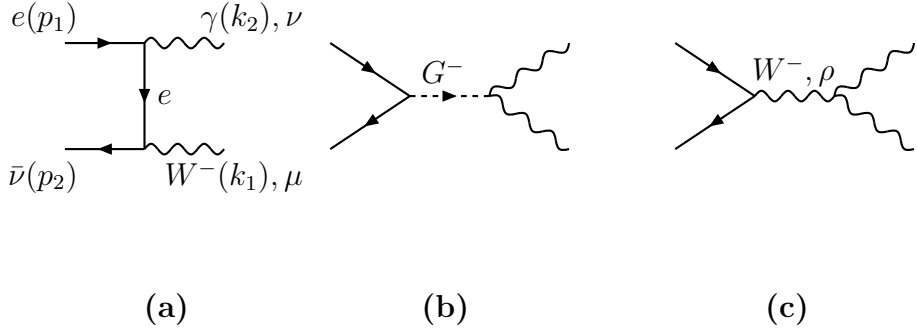
{\bf Fig.\ 4:} Amplitudes contributing to the reaction
$e^-\bar{\nu} \to W^-\gamma$ 
\end{center}

We set again ${\cal M}=\langle e^{-}\nu|T|e^{-}\nu\rangle$ and ${\cal T}=
\langle e^{-}\nu|T|W^{-}\gamma\rangle$. As in the case of QCD,  the amplitude
consists of two distinct parts, a part that contains an electron propagator
(Fig. 4(a)) and a part that does not, which is shown in Figs. 4(b) and 4(c). 
As before, we denote them by ${\cal T}_t$ and ${\cal T}_s(\xi_w )$, 
respectively. We first define 
\begin{equation}
V_L^\mu\ =\ \frac{g_w}{2\sqrt{2}}\,
\bar{v}(p_2)\gamma^{\mu}(1-\gamma_{5})u(p_1) 
\end{equation}
and 
\begin{equation}
S_R\ =\ \frac{g_w}{2\sqrt{2}}\, \frac{m_e}{M_{W}}\,
\bar{v}(p_2)(1+\gamma_{5})u(p_1)\, .
\end{equation}
Clearly, one has the EWI
\begin{equation}
q_{\mu}V_L^{\mu}\ =\ M_W S_R\, .
\label{VVWI}
\label{VV}
\end{equation}
The amplitude ${\cal T}_{s}$ can the be written down in the closed form
\begin{equation}
\label{Tsxiw}
{{\cal T}_{s}}_{\mu\nu}(\xi_{w})\ =\ i
V_L^\lambda\, \Delta_{0\lambda}^{(\xi_{w}),\rho}(q)\,
\Gamma^{\gamma W^-W^+}_{\nu\rho\mu}\ +\ iS_R\, 
D_0^{(\xi_{w})}(q)\, \Gamma^{\gamma G^- W^+} _{\nu\mu}\, ,
\end{equation}
where $\Gamma^{\gamma W^-W^+}_{\nu\rho\mu}=e \Gamma_{\nu\rho\mu}
(-k_2,q,-k_1)$ is the tree-level $\gamma W^- W^+$ vertex and $\Gamma^{\gamma
G^- W^+}_{\nu\mu} = eM_Wg_{\mu\nu}$ is the tree-level $\gamma G^-W^+$ vertex.
In the expression (\ref{Tsxiw}), we explicitly display the dependence on the
GFP $\xi_{w}$. In addition, the amplitude ${\cal T}_{t}$ is given by 
\begin{equation}
{\cal T}_{t}^{\mu\nu}\ =\ \frac{ieg_w}{2\sqrt{2}}\,
\bar{v}(p_2)\, \gamma^\mu (1-\gamma_{5})\, 
\frac{1}{\not\! p_1-\not\! k_2 -m_e}\,  \gamma^\nu\,  u(p_1)\, . 
\end{equation}
Notice that ${\cal T}_{t}^{\mu\nu}$ does not depend on $\xi_{w}$. Denoting by
$k_1$ the four-momentum of the $W$ and by $k_2$ that of the photon, Eq.\
(\ref{OTWg}) becomes
\begin{equation}
\label{Mew}
\Im m {\cal M}\ =\  
{\cal T}_{\mu\nu}Q^{\mu\rho}(k_1)P^{\nu\sigma}(k_2,\eta)
{\cal T}^*_{\rho\sigma}\, ,
\end{equation}
where $P^{\mu\nu}$ is the photon polarization tensor given in Eq.\ 
(\ref{PhotPol}), and
\begin{equation}
\label{WPol}
Q^{\mu\nu} (k)\ =\ -g^{\mu\nu}\, +\, \frac{k^\mu k^\nu}{M^2_W}\, 
\end{equation}
is the $W$ polarization tensor. The polarization tensor $Q^{\mu\nu} (k)$
shares the property that, on shell, {\em i.e.},  for $k^2= M^2_{W}$, $k^{\mu}
Q_{\mu\nu}(k)=0$. Furthermore, in Eq.\ (\ref{Mew}), we omit the integration
measure $1/2\int dX_{LIPS}$. 

First, we will show how the dependence on the GFP $\xi_{w}$ cancels. To that
end, we employ the usual decomposition 
\begin{equation}
\Delta_{0\mu\nu}^{(\xi_{w})}(q) \
=\  U_{\mu\nu}(q)\ -\ \frac{q_\mu q_\nu}{M_W^2} D_0^{(\xi_{w})}(q^2)\, ,
\label{D0xi}
\end{equation}
the EWI  
\begin{equation}
q^{\rho}\Gamma^{\gamma W^- W^+}_{\nu\rho\mu}(-k_2,q,-k_1)\, 
                           Q^{\mu\lambda}(k_1)P^{\nu\sigma}(k_2,\eta)\ =\ M_W
\Gamma^{\gamma G^- W^+}_{\mu\nu}\, Q^{\mu\lambda}(k_1)
P^{\nu\sigma}(k_2,\eta)
\end{equation}
and the EWI of Eq.\ (\ref{VVWI}), and we obtain the following 
$\xi_{w}$-independent expression for ${\cal T}_{s}^{\mu\nu}$ 
\begin{eqnarray}
\label{Tsinter}
{\cal T}_{s}^{\mu\nu}&=&
ieV_L^\lambda U_{\lambda\rho}(q)\Gamma^{\nu\rho\mu}(-k_2,q,-k_1)\ =\ 
ieV_L^{\lambda}\, U_{\lambda\rho}(q)\, 
\Gamma^{F,\nu\rho\mu}(-k_2,q,-k_1)\nonumber\\
&=& {{\cal T}_{s}}^{F,\mu\nu}\, ,
\end{eqnarray}
where contraction over the polarization tensors $Q_{\mu\nu}$ and $P_{\mu\nu}$
is implied. In the last step of Eq.\ (\ref{Tsinter}), we have used the fact
that the $\Gamma^{P}$ part of the vertex, defined in Eq.\ (\ref{GFGP}),
vanishes when contracted with the polarization tensors. 

Next, we show how the dependence on the four-vector $\eta_{\mu}$ and the
parameter $\eta^2$ vanishes. First, it is straightforward to verify the
following EWI:
\begin{eqnarray}
k^\mu_1 \Gamma_{\nu\rho\mu}^F(-k_2,q,-k_1) & = & 
[U^{-1}_\gamma (k_2) - U^{-1}(q)- U^{-1}(k_1)]_{\nu\rho}\nonumber\\
&&+2M^2_W g_{\nu\rho} + (k_1 - k_2 )_\nu {k_1}_\rho\nonumber\\ 
&=&-U_{\nu\rho}^{-1}(q)\, +\, 2M^2_W g_{\nu\rho}\, -\, 
k_{2\nu} (k_1 - k_2 )_\rho \, ,
\label{WIx}
\end{eqnarray}
where the on-shell conditions $k_1^2=M^2_W$ and $k_2^2=0$ are used in the last
equality of Eq.\ (\ref{WIx}). Similarly, one has
\begin{eqnarray}
k^\nu_2 \Gamma_{\nu\rho\mu}^F(-k_2,q,-k_1) & = & 
[U^{-1}(q)- U^{-1}(k_1) +U^{-1}_\gamma (k_2)]_{\rho\mu}\nonumber\\
&&+ k_{2\rho} (k_1 - k_2 )_\mu \nonumber\\
&=&U_{\rho\mu}^{-1}(q)\, -\, (k_1 - k_2 )_\rho k_{1\mu}\, ,
\label{WIy} 
\end{eqnarray}
with
\begin{eqnarray}
U_{\alpha\beta}^{-1}(q)& =& (q^2-M^2_W)\, t_{\alpha\beta}\ +\
M^2_W\, \ell_{\alpha\beta}\, ,\nonumber\\
{U^{-1}_\gamma}_{\alpha\beta}(q)&=& q^2\, t_{\alpha\beta}\, .
\end{eqnarray}
So, when the $\eta^{\sigma}k_2^{\nu}$ term from $P_{\nu\sigma}(k_2,\eta )$
gets contracted with ${\cal T}_{\mu\nu}$, we have 
\begin{eqnarray}
\eta^\sigma k_2^\nu {{\cal T}_{s}}_{\mu\nu} &=& 
i e\eta^\sigma V_L^{\lambda}\Big[ g_{\lambda\mu}\ -\ 
U^\alpha_\lambda(q)\, U^{-1}_{\alpha\mu} (k_1)\Big], \nonumber\\
\eta^{\sigma}k_2^{\nu}{{\cal T}_{t}}_{\mu\nu} & =& 
-i e\eta^{\sigma}V_{L\mu}\, .
\end{eqnarray}
Adding the last two equations by parts, we find
\begin{equation} 
\eta^\sigma k_2^\nu {\cal T}_{\mu\nu}\ =\  i e\eta^\sigma
V_{L}^\lambda\, U^\alpha_\lambda(q)\, U^{-1}_{\alpha\mu} (k_1)\, .
\end{equation}
Since the result is proportional to ${k_1}_{\mu}$, the four-momentum of the
external $W$ boson, we immediately see that 
\begin{equation}
\eta^{\sigma}k_2^{\nu}{\cal T}_{\mu\nu}Q^{\mu\rho}(k_1)\ =\ 0\, .
\end{equation}
For the same reasons, the term proportional to $\eta^2$ vanishes as well. 
Consequently, $\Im m {\cal M}$ takes on the form
\begin{eqnarray}
\Im m {\cal M}&=& -({\cal T}_{s}^{F}+{\cal T}_{t})_{\mu\nu}Q^{\mu\rho}(k_1)
( {\cal T}_{s}^{F}+{\cal T}_{t})^{*}_{\rho\nu}\nonumber\\
&=& ( {\cal T}_{s}^{F}+{\cal T}_{t})^{\mu\nu}
( {\cal T}_{s}^{F}+{\cal T}_{t})^{*}_{\mu\nu}\ -\ 
( {\cal T}_{s}^{F}+{\cal T}_{t})^{\mu\nu}
\, \frac{k_{1\mu} k_1^\rho }{M_W^2 }\, 
( {\cal T}_{s}^{F}+{\cal T}_{t})^{*}_{\rho\nu}\nonumber\\
&=&\Im m {\cal M}^{a}+\Im m {\cal M}^{b} .
\label{XYZ}
\end{eqnarray}
The absorptive sub-amplitude, $\Im m {\cal M}^{a}$, consists of three terms, 
\begin{eqnarray}
\Im m {\cal M}^{a} &=&{\cal T}_{s}^{F}{{\cal T}_{s}^{F}}^{*}+
( {\cal T}_{s}^{F}{\cal T}_{t}^{*}+  {\cal T}_{t} {{\cal T}_{s}^{F}}^{*})
+ {\cal T}_{t}{\cal T}_{t}^{*}\nonumber\\
&=& \Im m \widehat{{\cal M}}^{a}_1+
\Im m \widehat{{\cal M}}^{a}_2+\Im m \widehat{{\cal M}}^{a}_{3}\, .
\end{eqnarray}
The first term, $\Im m \widehat{{\cal M}}^{a}_1$, can easily be identified 
with a propagator-like contribution. In particular, using Eq.\ (\ref{FRE}), 
we find 
\begin{equation}
\Im m \widehat{{\cal M}}^{a}_1\ =\ e^2\, 
V_L^{\rho}\, U_{\rho\mu}(q)\, 
\Big[ -8 q^2t^{\mu\nu}(q)\, +\,
4{(k_1-k_2)}^{\mu}{(k_1-k_2)}^{\nu} \Big] U_{\nu\lambda}(q)\, 
                                                           V_L^{\lambda}\, .
\label{IMA1}
\end{equation}
The amplitudes, $\Im m \widehat{{\cal M}}^{a}_2$ and 
$\Im m \widehat{{\cal M}}^{a}_{3}$, are vertex- and box-like contributions, 
respectively, and they will not be considered any further here. 

We must now isolate the corresponding propagator-like piece from $\Im m {\cal
M}^{b}$. By virtue of the EWI of Eq.\ (\ref{WIx}), we have
\begin{equation}
k_1^{\mu}{{\cal T}_{s}}^{F}_{\mu\nu}\ =\ -ieV_{L\nu}\, -\, ie 
V_{L\lambda}\, U^{\lambda\rho}(q)\, \Big[ (k_1-k_2)_\rho {k_2}_\nu
- 2M^2_Wg_{\rho\nu} \Big]\, .
\label{A1}
\end{equation}
In addition, we evaluate the EWI
\begin{eqnarray}
k_1^{\mu}{{\cal T}_{t}}_{\mu\nu} &=& 
ieV_{L\nu}\, +\, M_W\, \frac{ieg_wm_e}{2\sqrt{2}M_W}\, 
\bar{v}(p_2)\, (1+\gamma_5)\, \frac{1}{\not\! p_1 -\not\! k_2-m_e}\, 
\gamma_\nu\,  u(p_1)
\nonumber\\ 
&=& ieV_{L\nu}\ +\ M_W{\cal L}_{\nu}\, ,
\label{Gbox}
\end{eqnarray}
which is shown diagrammatically in Fig.\ 5. 
\begin{center}
\begin{picture}(340,200)(0,0)
\SetWidth{0.8}

\ArrowLine(20,150)(50,150)\Text(25,160)[r]{$e (p_1)$}
\ArrowLine(50,150)(50,110)\Text(55,130)[l]{$e$}
\ArrowLine(50,110)(20,110)\Text(25,100)[r]{$\bar{\nu} (p_2)$}
\Photon(50,150)(80,150){2}{3}\Text(110,160)[r]{$\gamma (k_2),\nu$}
\Photon(50,110)(80,110){2}{3}\Text(110,100)[r]{$W^- (k_1),\mu$}
\Text(110,130)[r]{$\cdot\, k_1^\mu$}
\Text(50,50)[l]{{\bf (a)}}

\Text(120,130)[l]{$=$}

\ArrowLine(140,150)(170,130)
\ArrowLine(170,130)(140,110)
\Photon(170,130)(200,150){2}{3}\Text(205,160)[r]{$\gamma $} 
\Photon(170,130)(200,110){2}{3}\Text(200,100)[r]{$W^- $}
\Text(170,50)[l]{{\bf (b)}}

\Text(225,130)[l]{$+$}
\Text(240,130)[l]{$M_W\, \cdot$}

\ArrowLine(270,150)(300,150)
\ArrowLine(300,150)(300,110)\Text(305,130)[l]{$e$}
\ArrowLine(300,110)(270,110)
\Photon(300,150)(330,150){2}{3}\Text(340,160)[r]{$\gamma$}
\DashArrowLine(300,110)(330,110){3}\Text(340,100)[r]{$G^-$}
\Text(300,50)[l]{{\bf (c)}}

\end{picture}\\
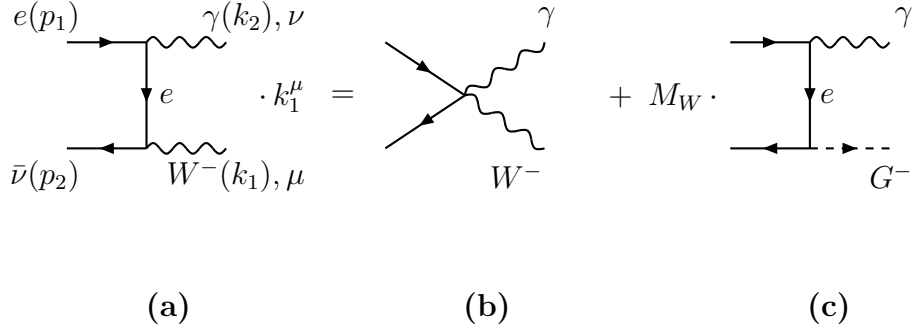
{\bf Fig.\ 5:} Elementary BRS identity for the $e$-dependent
amplitude ${\cal T}^{\mu\nu}_t$
\end{center}

\noindent
Adding Eqs.\ (\ref{A1}) and (\ref{Gbox}) by parts, we obtain
\begin{equation}
k_1^{\mu}\, ( {\cal T}_{s}^{F}+{\cal T}_{t})_{\mu\nu}\ =\
-ieV_{L\lambda}\, U^{\lambda\rho}(q)\, 
\Big[(k_1-k_2)_{\rho}{k_2}_{\nu} - 2M^2_{W}g_{\rho\nu}
 \Big]\ +\ M_W{\cal L}_{\nu}\, .\quad
\end{equation}
Making now use of the EWI of Eq.\ (\ref{VV}) and writing 
\begin{equation}
S_R\ =\ M_W V_{L\mu}\, U^{\mu\nu}(q)\, q_\nu 
\end{equation} 
yields the following WI for ${\cal L}_{\sigma}$:
\begin{equation}
k_2^\nu\, {\cal L}_\nu\ =\ -ie S_R \ =\
-ieM_W\, V_{L\alpha}\, U^{\alpha\beta}(q)\, q_{\beta}\, .
\end{equation}
We also use the following algebraic identity
\begin{equation}
q^{\mu}{(k_1-k_2)}^{\nu}\ =\ 
2k_2^{\mu}{(k_1-k_2)}^{\nu}+ {(k_1-k_2)}^{\mu}{(k_1-k_2)}^{\nu}\, .
\end{equation}
Taking the above relations into account, we eventually obtain
\begin{eqnarray}
\Im m {\cal M}^{b}&=& -e^2\, V_{L\rho}\, U^{\rho\mu}(q)\, 
\Big[ 4M^2_W g_{\mu\nu}+ 2 (k_1-k_2)_\mu(k_1-k_2)_\nu \Big] 
U^{\nu\lambda}(q)\, V_{L\lambda}\nonumber\\
&& -2ieM_W\, \Big[ V_{L\rho}\, U^{\rho\nu}(q)\, {\cal L}_{\nu}^{*} \,
-\, {\cal L}_{\nu}\, U^{\nu\lambda}(q)\, V_{L\lambda} \Big]\
-\ {\cal L}^{\nu}{\cal L}_{\nu}^{*}\nonumber\\
&=& \Im m \widehat{{\cal M}}^{b}_1 + 
\Im m \widehat{{\cal M}}^{b}_2 + 
\Im m \widehat{{\cal M}}^{b}_{3}\, .
\label{IMA2}
\end{eqnarray}
Adding the two propagator-like parts $\Im m \widehat{{\cal M}}^{a}_1$ and 
$\Im m \widehat{{\cal M}}^{b}_1$  from Eqs.\ (\ref{IMA1}) and (\ref{IMA2}), 
respectively, we find
\begin{eqnarray}
\Im m \widehat{{\cal M}}_1&=& 
\Im m \widehat{{\cal M}}^{a}_1+ \Im m \widehat{{\cal M}}^{b}_1\nonumber \\
&=& e^2\, V_L^\rho\, U_{\rho\mu}(q)
\Big[ -8q^2t^{\mu\nu}(q)\, -\, 4M_{W}^2g^{\mu\nu}
+\, 2{(k_1-k_2)}^{\mu}{(k_1-k_2)}^{\nu}\Big]U_{\nu\lambda}(q)\, V_L^{\lambda}.
\nonumber\\
&&
\label{IMA}
\end{eqnarray}  
Next, we carry out the phase-space integration over $1/2\int dX_{LIPS}$, 
using the formulas given in Eqs.\ (\ref{LIPS1}) and (\ref{LIPS2}), and 
the fact that $\lambda^{1/2} (q^2,M_{W}^2,0)=q^2-M_{W}^2>0$. In this
way, we have
\begin{equation}
\Im m \widehat{{\cal M}}_1\ =\
V_{L\rho}U^{\rho\mu}(q)\ \Im m \widehat{\Pi}_{\mu\nu}^{W}\
U^{\nu\lambda}(q)\, V_{L\lambda}\, ,
\end{equation}
with
\begin{eqnarray}
\Im m \widehat{\Pi}_{\mu\nu}^{W}(q) &=& \Im m \widehat{\Pi}_T^{W}(q^2)\,
t_{\mu\nu}(q)\ +\ \Im m \widehat{\Pi}_L^{W}(q^2)\, 
\ell_{\mu\nu}(q),\nonumber\\
\Im m \widehat{\Pi}_T^{W}(q^2) &=& \frac{\alpha_{em}}{2}\, (q^2-M^2_W)\,
\Big( -\frac{11}{3}\, +\, \frac{4M^2_W}{3q^2}\, +\, \frac{M^4_W}{3q^4}
\Big)\, ,\nonumber\\
\Im m \widehat{\Pi}_L^{W}(q^2) &=& \frac{\alpha_{em}}{2}\, (q^2-M^2_W)\,
\Big( -\, \frac{2M^2_W}{q^2}\, +\, \frac{M^4_W}{q^4}
\Big)\, .
\end{eqnarray}
Here, $\alpha_{em}=e^2/(4\pi)$ is the electromagnetic fine structure constant.
The real part of the transverse, on-shell renormalized, $W$-boson self-energy,
$\Re e\widehat{\Pi}_T^{W,R}(s)$, can be determined by means of a doubly
subtracted DR given in Eq.\ (\ref{DR3}). Furthermore, we have to assume a
fictitious photon mass, $\mu_\gamma$, in order to regulate the infra-red (IR) 
divergences. More explicitly, the DR of our interest reads
\begin{eqnarray}
\label{RDRW}
\Re e\widehat{\Pi}_T^{W,R}(s) &=& \Re e\widehat{\Pi}_T^W (s)\, -\,
(s-M^2_W)\Re e\widehat{\Pi}_T^W{}'(M^2_W)\, -\, \Re e\widehat{\Pi}_T^W(M^2_W)
\nonumber\\
&=& \lim\limits_{\Lambda\to \infty}\lim\limits_{\mu_\gamma\to 0}
\frac{(s-M^2_W)^2}{\pi}\ \mbox{P}\, \int\limits^\Lambda_{(M_W+\mu_\gamma )^2}
\, \frac{ ds'\, \Im m \widehat{\Pi}^W_T(s')}{(s'-M^2_W)^2 (s'-s)}\, .
\end{eqnarray}
To obtain the analytic form of $\Re e\widehat{\Pi}_T^{W,R}(s)$, we first
evaluate the following integrals:
\begin{eqnarray}
\label{F0}
F_0(s) &=& (s-M^2_W)\ \mbox{P}\, \int\limits^\infty_{(M_W+\mu_\gamma )^2}
\, ds'\, \frac{1}{(s'-M^2_W) (s'-s)}\nonumber\\
&=&-\, \ln\Big(\, \frac{|s-M^2_W|}{2M_W\mu_\gamma}\, \Big)\, ,\\[0.3cm]
\label{F1}
F_1(s) &=& (s-M^2_W)\ \mbox{P}\, \int\limits^\infty_{(M_W+\mu_\gamma )^2}
\, ds'\, \frac{1}{(s'-M^2_W) (s'-s)}\, \frac{M^2_W}{s'}\nonumber\\
&=&-\, \frac{M^2_W}{s}\, \ln\Big(\, \frac{|s-M^2_W|}{2M_W\mu_\gamma}\, \Big)\,
-\, \Big(1-\frac{M^2_W}{s}\Big)\ln\Big(\frac{M_W}{2\mu_\gamma}\Big)\, ,
\\[0.3cm]
\label{F2}
F_2(s) &=& (s-M^2_W)\ \mbox{P}\, \int\limits^\infty_{(M_W+\mu_\gamma )^2}
\, ds'\, \frac{1}{(s'-M^2_W) (s'-s)}\, \frac{M^4_W}{s'^2}\nonumber\\
&=&-\, \frac{M^4_W}{s^2}\, \ln\Big(\, \frac{|s-M^2_W|}{2M_W\mu_\gamma}\, 
\Big)\, -\, \ln\Big(\frac{M_W}{2\mu_\gamma}\Big)\, +\, 1\, -\, 
\frac{M^2_W}{s}\, ,
\end{eqnarray}
Armed with the integrals defined in Eqs.\ (\ref{F0})--(\ref{F2}), one then
obtains
\begin{equation}
\label{RenWPT}
\Re e\widehat{\Pi}_T^{W}(s)\ =\ \frac{\alpha_{em}}{2}\, (s-M^2_W)\,
\Big( -\frac{11}{3}F_0\, +\, \frac{4}{3}F_1\, +\, \frac{1}{3}F_2 \Big)\, .
\end{equation}
Eq.\ (\ref{RenWPT}) coincides with the PT $W$-boson self-energy \cite{JPself}
or equivalently with the $W$-boson self-energy computed in the BFG for 
$\xi_Q=1$ \cite{DDW}.
 
\setcounter{equation}{0}
\section{Cutkosky considerations}
\indent

In this section, we focus on the LHS of the OT and present a different point
of view and a self-consistency check. In particular, we consider the one-loop
$S$-matrix element for a given process and compute its imaginary part by direct
application of the Cutkosky rules. The expressions so obtained consist of the
product of tree-level amplitudes, with the important difference that now
``unphysical'' degrees of freedom appear as intermediate states, giving in
turn rise to ``unphysical'' thresholds. These tree-level amplitudes are
related by EWIs. We show that, when fully exploited, these EWIs give rise to
propagator-, vertex- and box-like expressions, which contain physical 
thresholds only, whereas all the unphysical thresholds disappear completely.
The expressions so derived are identical to the imaginary parts of the 
corresponding PT Green's functions, which one can obtain directly from the 
$S$-matrix. Also, both real and imaginary parts are related via a DR, as
has been discussed in Section 2. 

For the process $l\nu_l\to W^- (p) H(p_H)$, we have in
an arbitrary $\xi$ gauge
\begin{eqnarray}
\frac{p^\mu}{M_W}\, T^{(\xi )}_{(a)\, \mu} & =&
T^{(\xi )}_{(b)}\ +\ \frac{ig_w}{2M_W}S_R, \label{WH1}\\
\frac{p^\mu}{M_W}\, T^{(\xi )}_{(c)\, \mu} & =& 
T^{(\xi )}_{(d)}\ -\ \frac{ig_w}{2M_W}S_R\, . \label{WH2}
\end{eqnarray}
\begin{center}
\begin{picture}(340,480)(0,0)
\SetWidth{0.8}

\ArrowLine(20,450)(60,420)\Text(25,460)[r]{$e$}
\ArrowLine(60,420)(20,390)\Text(25,380)[r]{$\bar{\nu}$}
\Photon(60,420)(100,420){2}{4}\Text(70,435)[l]{$W^-$}
\DashArrowLine(100,420)(140,450){3}\Text(150,460)[r]{$H(p_H)$}
\Photon(100,420)(140,390){2}{4}\Text(150,380)[r]{$W^-(p),\mu$}

\ArrowLine(220,450)(260,420)\Text(225,460)[r]{$e$}
\ArrowLine(260,420)(220,390)\Text(225,380)[r]{$\bar{\nu}$}
\DashArrowLine(260,420)(300,420){3}\Text(270,435)[l]{$G^-$}
\DashArrowLine(300,420)(340,450){3}\Text(340,460)[r]{$H$}
\Photon(300,420)(340,390){2}{4}\Text(340,380)[r]{$W^-,\mu$}

\Text(155,350)[l]{{\bf (a)}}

\ArrowLine(20,300)(60,270)\Text(25,310)[r]{$e$}
\ArrowLine(60,270)(20,240)\Text(25,230)[r]{$\bar{\nu}$}
\Photon(60,270)(100,270){2}{4}\Text(70,285)[l]{$W^-$}
\DashArrowLine(100,270)(140,300){3}\Text(140,310)[r]{$H$}
\DashArrowLine(100,270)(140,240){3}\Text(140,230)[r]{$G^-$}

\ArrowLine(220,300)(260,270)\Text(225,310)[r]{$e$}
\ArrowLine(260,270)(220,240)\Text(225,230)[r]{$\bar{\nu}$}
\DashArrowLine(260,270)(300,270){3}\Text(270,285)[l]{$G^-$}
\DashArrowLine(300,270)(340,300){3}\Text(340,310)[r]{$H$}
\DashArrowLine(300,270)(340,240){3}\Text(340,230)[r]{$G^-$}

\Text(155,200)[l]{{\bf (b)}}

\ArrowLine(20,150)(60,150)\Text(25,160)[r]{$e$}
\ArrowLine(60,150)(60,100)\Text(65,125)[l]{$e$}
\ArrowLine(60,100)(20,100)\Text(25,90)[r]{$\bar{\nu}$}
\DashArrowLine(60,150)(100,150){3}\Text(100,160)[r]{$H$}
\Photon(60,100)(100,100){2}{3}\Text(110,90)[r]{$W^-,\mu $}
\Text(55,40)[l]{{\bf (c)}}

\ArrowLine(220,150)(260,150)\Text(225,160)[r]{$e$}
\ArrowLine(260,150)(260,100)\Text(265,125)[l]{$e$}
\ArrowLine(260,100)(220,100)\Text(225,90)[r]{$\bar{\nu}$}
\DashArrowLine(260,150)(300,150){3}\Text(300,160)[r]{$H$}
\DashArrowLine(260,100)(300,100){3}\Text(300,90)[r]{$G^-$}
\Text(255,40)[l]{{\bf (d)}}

\end{picture}\\
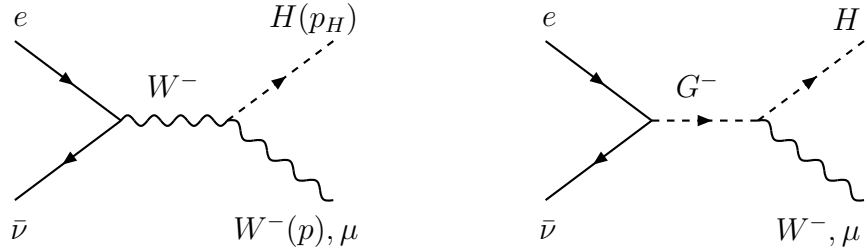
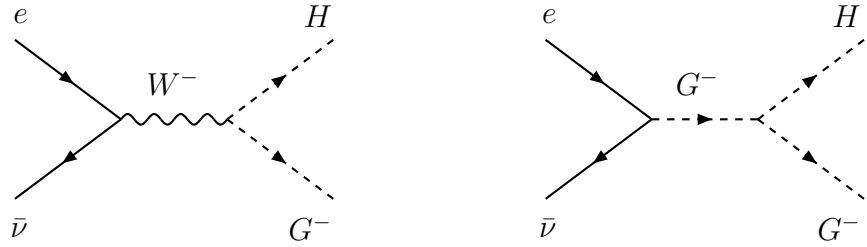
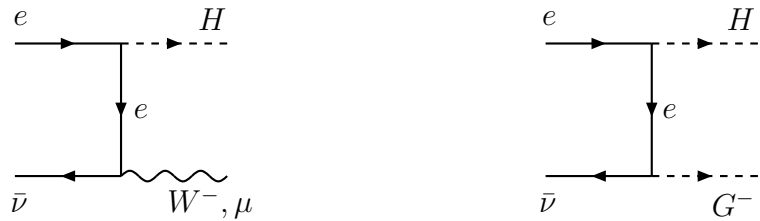
{\bf Fig.\ 6:} Graphs contributing to the amplitudes $T_{(a)\mu}^{(\xi)}$,
$T_{(b)}^{(\xi )}$, $T_{(c)}$, and $T_{(d)}$.
    
\end{center}
We will carry out an explicit calculation of the $\Im m \widehat{\cal M}_1$ of
the process $e\bar{\nu}_e\to e\bar{\nu}_e$ at the one-loop electroweak order,
working on the LHS of the OT.  To simplify the algebra, we will assume that
only the $W$ and $H$ particles can come kinematically on the mass shell, as
shown in Fig.\ 6. In what follows, we omit the common integration measure of
the loop, $1/[2(2\pi)^4]\int d^4p d^4p_H \delta^{(4)}(p_H +p -p_e -p_\nu )$.
Then, the absorptive amplitude, $\Im m{\cal M}$, for the aforementioned
process may be written as (suppressing contraction over Lorentz indices, and
using the on-shell conditions $p_{H}^2=M_{H}^2$, $p^2=M_{W}^2$) 
\begin{eqnarray}
\Im m{\cal M} & = & \tilde{\Delta}_{0H}(p_H)\Big[\,
T^{(\xi)}_{(a)} \tilde{\Delta}_0^{(\xi)}(p) T^{(\xi)*}_{(a)}
\, +\, T^{(\xi)}_{(b)} \tilde{D}_0^{(\xi)}(p) T^{(\xi)*}_{(b)}
+\, T^{(\xi)}_{(c)} \tilde{\Delta}_0^{(\xi)}(p) T^{(\xi)*}_{(a)}\nonumber\\
&& +\, T^{(\xi)}_{(a)} \tilde{\Delta}_0^{(\xi)}(p) T^{(\xi)*}_{(c)}
\, +\, T^{(\xi)}_{(d)} \tilde{D}_0^{(\xi)}(p) T^{(\xi)*}_{(b)}
\, +\, T^{(\xi)}_{(b)} \tilde{D}_0^{(\xi)}(p) T^{(\xi)*}_{(d)}\nonumber\\
&& +\, T^{(\xi)}_{(c)} \tilde{\Delta}_0^{(\xi)}(p) T^{(\xi)*}_{(c)}
\, +\, T^{(\xi)}_{(d)} \tilde{D}_0^{(\xi)}(p) T^{(\xi)*}_{(d)}\, \Big]\, ,
\label{ImM}
\end{eqnarray}
where the tilde acting on the tree-level propagators simply projects out the
corresponding absorptive parts. Such a projection can effectively be obtained
by applying the Cutkosky rules. More explicitly, we have
\begin{eqnarray}
\tilde{\Delta}_{0H} (p_H) & = & 2\pi\, \delta_+ (p^2_H-M^2_H)\, ,\\
\tilde{D}_0^{(\xi)} (p) & =& 2\pi\, \delta_+ (p^2 - \xi M^2_W)\, ,\\
\tilde{\Delta}_{0\, \mu\nu}^{(\xi)} (p) & = & 2\pi\, \Big[
Q_{\mu\nu}(p)\, \delta_+ (p^2 -M^2_W)\,
-\, \frac{p_\mu p_\nu}{M^2_W} \delta_+ (p^2 -\xi M^2_W)\Big]\nonumber\\
&=& \tilde{U}_{\mu\nu}(p)\ -\ \frac{p_\mu p_\nu}{M^2_W}\tilde{D}_0^{(\xi)}(p)
\, ,
\end{eqnarray}
where the $W$-boson polarization tensor $Q_{\mu\nu}(p)$ is given in Eq.\
(\ref{WPol}) and $\delta_+ (p^2-M^2) = \delta (p^2 - M^2)\theta (p^0)$.  After
identifying the PT piece, $T_P=-ig_wS_R/(2M_W)$, which is obtained from
Eq.~(\ref{WH2}) each time the $p^\mu p^\nu$-dependent part of
$\tilde{\Delta}^{(\xi)}_{0\mu\nu}$ gets contracted with $T^{(\xi)}_{(c)}$, we
observe that the imaginary propagator-like part may be decomposed as follows:
\begin{equation}
\Im m \widehat{\cal M}_1 = \Im m \widehat{\cal M}_1^{(phys)}
+ \delta\widehat{\cal M}_1\, , 
\end{equation}
where
\begin{eqnarray}
\Im m \widehat{\cal M}_1^{(phys)} & = & 
\tilde{\Delta}_{0H}(p_H)(2\pi)\delta_+ (p^2-M^2_W )\, \Big( 
T^{(\xi)}_{(a)\mu}Q^{\mu\nu}(p) T^{(\xi)*}_{(a)\nu}+
T_{P}\frac{p^\nu}{M_W}T^{(\xi)*}_{(a)\, \nu} \nonumber\\
&& + T^{(\xi )}_{(a)\, \lambda} \frac{p^\lambda}{M_W} T^*_{P}
+ T_{P} T^*_{P}\,\Big)
\label{dT1}
\end{eqnarray} 
and 
\begin{eqnarray}
\delta\widehat{\cal M}_1 
& = & -\, \tilde{\Delta}_{0H}(p_H)\tilde{D}_0^{(\xi)}(p)
\Big( T^{(\xi)*}_{(a)\, \lambda} \frac{p^\lambda p^\nu}{M^2_W}
T^{(\xi)*}_{(a)\, \nu}
-T^{(\xi)}_{(b)} T^{(\xi)*}_{(b)} + 
T_{P}\frac{p^\nu}{M_W}T^{(\xi)*}_{(a)\, \nu}
\nonumber\\
&& + T^{(\xi )}_{(a)\, \lambda} \frac{p^\lambda}{M_W} T^*_{P}
+ T_{P} T^*_{P}\Big)\, .
\end{eqnarray} 
In the first term, $\Im m \widehat{\cal M}_1^{(phys)}$, we have collected all
contributions originating from the physical poles at $p^2_H=M^2_H$ and
$p^2=M^2_W$, whereas all those occuring at $p^2=\xi M^2_W$ and are
proportional to $\tilde{D}_0^{(\xi)}(p)$ are included in $\delta\widehat{\cal
M}_1$. 

The first important observation is that $\delta\widehat{\cal M}_1=0$, which
can be shown with the help of the EWI in Eq.~(\ref{WH1}). So, the full 
exploitation of this WI gives rise to a propagator-like imaginary part
where all unphysical thresholds have been cancelled. In addition, with the
help of the same WI, we obtain for $\Im m \widehat{\cal M}_1^{(phys)}$, 
\begin{equation}
\Im m \widehat{\cal M}_1\ =\ \Im m \widehat{\cal M}_1^{(phys)}\ =\
\frac{1}{2}\, \int dX_{LIPS}\, \Big( 
-T^{(\xi)}_{(a)} T^{(\xi)*}_{(a)}+
T^{(\xi)}_{(b)} T^{(\xi)*}_{(b)} \Big)\, .
\end{equation}
We must now demonstrate that the final dependence on $\xi$ cancels in the
above equation.  Notice that even though we use the on shell conditions
$p^2=M^2_{W}$ and $p_{H}^2=M_{H}^2$, the amplitudes $T$ in the last equation
are {\em not} really ``on shell'', because they are {\em not} contracted by
the corresponding polarization vectors; therefore the $\xi$-cancellation is
not immediate.  To verify the cancellation, we must employ the identity of
Eq.\ (\ref{D0xi}) to decompose the internal tree-level $W$ propagators, and
the WIs, which relate the tree-level vertices involved, {\em i.e.}, 
\begin{eqnarray}
q^{\nu} \Gamma^{HW^+W^-}_{0\mu\nu} &= & - M_W \Gamma^{HW^+G^-}_{0\mu}\, +\,
\frac{ig_w}{2}\, M_W p_\mu\, ,\nonumber\\
q^{\nu} \Gamma^{HG^+W^-}_{0\nu} &= & - M_W \Gamma_0^{HG^+G^-}\, -\,
\frac{ig_w}{2}\, M_W^2\, . 
\end{eqnarray}
Thus, the final expression can be cast into the form
\begin{equation}
\Im m \widehat{\cal M}_1\ =\ \frac{1}{2}\, \int dX_{LIPS}\, \Big(
-T^{(\infty)}_{(a_1)} T^{(\infty)*}_{(a_1)}+
T^{(\infty)}_{(b_1)} T^{(\infty)*}_{(b_1)}\Big)\, ,
\label{aaa}
\end{equation}
where by the index $a_1$ ($b_1$) denotes the first graph in Fig.\ 6a (6b), and
the superscript ``$\infty$'' means that the internal tree-level $W$
propagators are in the unitary gauge. 

This is precisely what one would obtain from the straightforward computation
of the imaginary part of the one-loop PT $WW$ self-energy, presented in
\cite{JPself}. The expression for the GFP-independent propagator-like part of
$\widehat{\cal M}$, $\widehat{\cal M}_1$, in terms of the PT $WW$ self-energy,
$\widehat{\Pi}_{\mu\nu}(q)$, is given by 
\begin{equation}
\widehat{\cal M}_1\ =\ V_{L\sigma}
U^{\sigma\mu}(q)\, \widehat{\Pi}_{\mu\nu}(q)\, U^{\nu\rho}(q)\, 
V_{L\rho}\, .
\end{equation}
The Higgs-dependent part of $\widehat{\Pi}_{\mu\nu}$, call it
$\widehat{\Pi}_{\mu\nu}^{(HW)}$, is  given by \cite{Pap}
\begin{equation}
\widehat{\Pi}_{\mu\nu}^{(HW)}(q)\ =\  \pi\alpha_w\,
\int \frac{d^nk}{i(2\pi)^n}\, I(q,k)\,
[(2k+q)_{\mu}(2k+q)_{\nu}-4 M^2_{W}g_{\mu\nu}]\, ,
\label{oldstuff}
\end{equation}
where $\alpha_w=g^2_w/(4\pi)$ is the SU$(2)_L$ fine structure constant and
\begin{equation}
I(q,k)\ =\ \frac{1}{(k^2-M_W^2)[(k+q)^2-M_H^2]}\ .
\end{equation}
It is now easy to see that the imaginary part of
$\widehat{\Pi}_{\mu\nu}^{(HW)}$ is indeed equal to Eq.\ (\ref{aaa}). This can
be verified by an explicit application of the Cutkosky rules on the expression
in the RHS of Eq.\ (\ref{oldstuff}). Actually, this amounts to determining
where the logarithmic terms, which are obtained after the integration over the
virtual momenta, turn negative. One could then compare that result with the
result we will obtain after integrating Eq.\ (\ref{WH1}) over the phase space
integral given above. To that end, we must make use of the fact that the
typical integral over the Feynman parameter $x$ 
\begin{eqnarray}
\Im m\Big[\int \frac{d^nk}{i(2\pi)^n}\, I(q,k)\Big] &=&
-\frac{1}{16\pi^2}\, \Im m\Big\{ \, \int^1_{0} dx
\ln[M_{H}^2x+M_{W}^2(1-x)-q^2x(1-x)]\Big\}
\nonumber\\
&=& \frac{\theta [ q^2-(M_W+M_H)^2 ]}{8\pi q^2}\, 
\lambda^{1/2}(q^2,M^2_H,M^2_W)\nonumber\\
&=& \frac{1}{2}\int dX_{LIPS}\, .\qquad
\label{imhz}
\end{eqnarray}
The above relation  gives an explicit connection between Cutkosky rules and
the two-body LIPS given in Eq.\ (\ref{LIPS1}). As has been discussed in
Section 2, the analytic continuation of the logarithmic function in the RHS of
Eq.\ (\ref{imhz}) is uniquely determined via the prescription $s\to
s+i\varepsilon$.

It is important to emphasize the conclusions of this section: We have
proceeded in two different ways. First, we have calculated the propagator-like
imaginary part by applying the Cutkosky rule, and  exploiting the tree-level
EWIs. Then, we have computed the imaginary part of the one-loop PT $W$
self-energy, obtained by the usual $S$-matrix PT rules. The two analytic
results have turned out to be identical. We can therefore conclude that the PT
Green's functions, contrary to their conventional counterparts, satisfy
individually the OT. We consider that a crucial point for the success of our
resummation algorithm. In addition, the above analysis demonstrates that one
can work freely on either side of the OT and arrive at a unique result, just
by following the same rules, {\em i.e.}, by fully exploiting the EWIs of the
theory.

\setcounter{equation}{0}
\section{The Background Field Gauge}
\indent

The formulation of non-Abelian gauge field theories in the framework of the
BFG endows the $n$-point functions obtained from the generating functional
with a number of characteristic properties. Most remarkably, the BFG $n$-point
functions satisfy {\em tree-level} Ward identities, to all orders in
perturbation theory.  This fact is to be contrasted with the Slavnov-Taylor
identities of the conventional covariant formulation, where the tree-level WI
are spoiled by the appearance of ``ghost'' Green's function, as soon as
quantum corrections are introduced. On the other hand, the BFG $n$-point
functions display in general a residual dependence on the quantum GFP 
$\xi_{Q}$, which is used to ``gauge-fix'' the gauge fields inside the quantum
loops. As we will show in this section, the functional dependence of the BFG
two-point functions on $\xi_Q$ is such that it leads to the appearance of {\em
unphysical thresholds}, at $q^2=\xi_{Q} M^2$. 

What is rather striking in this context is the following observation. 
Consider a BFG two-point function computed at one-loop at some arbitrary
$\xi_{Q}$. Let us then separate it into two parts: the part that has only
physical thresholds (at $q^2=M^2$) and the part that has unphysical thresholds
(at $q^2=\xi_{Q} M^2$). Interestingly enough, one finds that each part 
satisfies {\em separately} the correct tree-level WI.
\begin{center}
\begin{picture}(340,350)(0,0)
\SetWidth{0.8}

\Photon(0,300)(30,300){2}{3}\Text(10,310)[r]{$\widehat{W}^+$}
\Photon(90,300)(120,300){2}{3}\Text(135,310)[r]{$\widehat{W}^+$}
\PhotonArc(60,300)(30,0,180){2}{8}\Text(60,340)[c]{$W^+$}
\DashCArc(60,300)(30,180,360){4}\Text(60,260)[c]{$H$}
\Text(60,220)[c]{{\bf (a)}}

\Photon(200,300)(230,300){2}{3}\Text(210,310)[r]{$\widehat{W}^+$}
\Photon(290,300)(320,300){2}{3}\Text(335,310)[r]{$\widehat{W}^+$}
\DashCArc(260,300)(30,0,180){4}\Text(260,340)[c]{$G^+$}
\DashCArc(260,300)(30,180,360){4}\Text(260,260)[c]{$H$}
\Text(260,220)[c]{{\bf (b)}}

\DashArrowLine(0,100)(30,100){4}\Text(10,110)[r]{$\widehat{G}^+$}
\DashArrowLine(90,100)(120,100){4}\Text(135,110)[r]{$\widehat{G}^+$}
\PhotonArc(60,100)(30,0,180){2}{8}\Text(60,140)[c]{$W^+$}
\DashCArc(60,100)(30,180,360){4}\Text(60,60)[c]{$H$}
\Text(60,20)[c]{{\bf (c)}}

\DashArrowLine(200,100)(230,100){4}\Text(210,110)[r]{$\widehat{G}^+$}
\DashArrowLine(290,100)(320,100){4}\Text(335,110)[r]{$\widehat{G}^+$}
\DashCArc(260,100)(30,0,180){4}\Text(260,140)[c]{$G^+$}
\DashCArc(260,100)(30,180,360){4}\Text(260,60)[c]{$H$}
\Text(260,20)[c]{{\bf (d)}}

\end{picture}\\
{\bf Fig.\ 7:} $WH$ contributions to 
$\Pi^{ \widehat{W}^+ \widehat{W}^+}_{\mu\nu}$ [(a),(b)]
and  $\Pi^{ \widehat{G}^+ \widehat{G}^+ }_{\mu\nu}$ [(c),(d)].
    
\end{center}
Defining $I_Q$ as follows:
\begin{equation}
I_{Q}(q,k)\ =\ \frac{1}{(k^2-\xi_Q M_{W}^2)[(k+q)^2-M_{H}^2]}
\end{equation}
and using the identity
\begin{equation}
\frac{1-\xi_{Q}}{(k^2-M_{W}^2)(k^2-\xi_Q M_{W}^2)}\ =\ 
\frac{1}{M_W^2}\, \Big[\, \frac{1}{k^2-M_{W}^2}\,
                                  -\, \frac{1}{k^2-\xi_Q M_{W}^2}\, \Big] ,
\end{equation}
we have for the Feynman diagrams (a) and (b) in Fig.\ 7 (loop integration,
$\int d^n k/i(2\pi)^n$, implied)
\begin{eqnarray}
\mbox{(a)} &=& g^2_w M_W^2\Big[\Big(-g_{\mu\nu}+
\frac{k_\mu k_\nu}{M^2_W}\Big)\, I(q,k)\ -\ 
\frac{k_\mu k_\nu}{M^2_W}\, I_Q (q,k)\Big],\nonumber\\
\mbox{(b)} &=& \frac{g^2_w}{4}(2k+q)_{\mu}(2k+q)_{\nu}\, I_{Q}(q,k)\, ,
\end{eqnarray}
from which follows that
\begin{eqnarray}
\Pi_{\mu\nu}^{(HW)}(q) &=& g^2_wM_W^2\Big[ \Big(-g_{\mu\nu}+
\frac{k_{\mu}k_{\nu}}{M^2_W}\Big)I(q,k)\ +\ 
\frac{1}{4M^2_W}\Big( (2k+q)_{\mu}(2k+q)_{\nu}\nonumber\\ 
&&-\, 4k_{\mu}k_{\nu} \Big) I_Q(q,k)\Big]\nonumber\\
&=& \bar{\Pi}_{\mu\nu}(q)\, +\, \Pi_{\mu\nu}^{Q}(q)\, ,
\end{eqnarray}
where $\bar{\Pi}_{\mu\nu}$ contains only physical thresholds, at
$q^2=(M_{W}+M_{H})^2$, whereas $\Pi_{\mu\nu}^{Q}$ contains unphysical
thresholds at $q^2=(\sqrt{\xi_Q} M_{W}+M_{H})^2$.  Similarly, from Figs.\ 7(c)
and 7(d), we calculate
\begin{eqnarray}
\mbox{(c)} &=& g^2_wq^{\rho}q^{\sigma}\Big[ \Big( 
          - g_{\rho\sigma}+\frac{k_{\rho}k_{\sigma}}{M^2_W}\Big) I(q,k)\ -\ 
                                        \frac{k_{\rho}k_{\sigma}}{M^2_W}\, 
                                                         I_{Q}(q,k)\Big]\, ,\\
\mbox{(d)} &=& \frac{g^2_w}{4M^2_W}(M_H^2-\xi_Q M_W^2 )^2 I_{Q}(q,k)\, ,
\end{eqnarray}
and so
\begin{eqnarray}
\Omega^{(HW)}(q) &=& g^2_w\Big[\, \frac{(qk)^2}{M_W^2}-q^2\Big]\, I(q,k)\
+\ g^2_w\Big[ \frac{(M^2_H-\xi_Q M^2_W)^2}{4M^2_W}\, -\, 
\frac{(qk)^2}{M^2_W}\Big]\, I_{Q}(q,k) 
\nonumber\\
&=& \bar{\Omega}(q)\ +\ \Omega^{Q}(q)\, .
\end{eqnarray}
It is elementary to check that up to irrelevant tadpole terms, the following
WIs hold: 
\begin{equation}
\label{BFGph}
q^{\mu}q^{\nu}\bar{\Pi}_{\mu\nu}(q)\ -\ M^2_W\bar{\Omega}(q)\ =\ 0
\end{equation}
and
\begin{equation}
\label{BFGun}
q^{\mu}q^{\nu}\Pi_{\mu\nu}^Q (q)\ -\ M^2_W\Omega^Q (q)\ =\ 0 \, .
\end{equation}
It is worth noticing that the tree-level Ward identities, Eqs.\ (\ref{BFGph}) and
(\ref{BFGun}), are {\em individually} satisfied by the contributions having
physical and gauge-dependent unphysical thresholds, respectively.  This
property is not an accidental feature of the specific example considered
above, but, as we will argue in a moment, it must be valid for any individual
contribution to an {\em analytic} two-point correlation function. On the other
hand, it is obvious that neither $\bar{\Pi}$ nor $\Pi^Q$ can be obtained from
a specific choice of the $\xi_Q$ value. An exception to this is the value
$\xi_Q=1$. In this gauge, the physical and unphysical sectors are not
distinguishable.  If we impose the constraint of the absence of unphysical
thresholds in the BFG ---a property which is always preserved within the PT
framework \cite{JP&AP}, then the two-point correlation functions of the PT and
the BFG for $\xi_Q=1$ have to coincide at one loop. This feature should also 
hold true for all $n$-point functions at one loop. 

In the following, we argue that the reason which forces
$\bar{\Pi}_{\mu\nu}(q)$ and $\Pi^Q_{\mu\nu} (q)$ to satisfy individually the
same tree-like Ward identities as those of the full $\Pi_{\mu\nu}(q)$, is the
analyticity of $\Pi_{\mu\nu}(q)$. In fact, it is sufficient to show that $\Im
m \Pi_{\mu\nu}(q)=\Im m\bar{\Pi}_{\mu\nu}(q)\not= 0$ for a finite domain of
$q^2$ (for $\xi_Q\ne 1$). Then, Eq.\ (\ref{BFGph}) will be valid for the
finite kinematic region and will also hold true for any $q^2$, since $\Im
m\bar{\Pi}_{\mu\nu}$ is analytic.  That $\Re e\bar{\Pi}_{\mu\nu}$ will also
satisfy Eq.\ (\ref{BFGph}) is guaranteed through a DR. Finally, it is evident
that $\Pi^Q_{\mu\nu}(q)= \Pi_{\mu\nu}(q) - \bar{\Pi}_{\mu\nu}(q)$ will obey
the same WI (\ref{BFGun}). 

To give a specific example, let us consider the absorptive part of the $WW$
self-energy in the BFG at one loop, in which only the $W\gamma$ contributions
are considered. It is clear that, for the finite domain $M^2_W<q^2 <
\mbox{min}[\sqrt{\xi_Q}M^2_W, (M_W+M_Z)^2]$ ($\xi_Q\ne 1$), $\Im m
\Pi_{\mu\nu}(q)=\Im m\bar{\Pi}^{(\gamma W)}_{\mu\nu}(q)$. The latter leads to
the fact that $\bar{\Pi}^{(\gamma W)}_{\mu\nu}(q)$ satisfies Eq.\
(\ref{BFGph}) independently, for any $q^2$.  Similar arguments can carry over
to the other distinct threshold contributions. 

\setcounter{equation}{0}
\section{Issues of uniqueness}
\indent  

In this section, we will address issues related to the uniqueness of the PT
rearrangement. We know that the PT rearrangement gives rise to effective
self-energies ($\widehat{\Pi}$), vertices ($\widehat{\Gamma}$) and box graphs
($\widehat{B}$), endowed with several characteristic properties. The question 
naturally arises whether these effective Green's functions are unique. By
``unique'' we mean, whether after the PT rearrangement has been completed, one
could still define new Green's functions, by moving GFP-independent
terms around, in such a way as: 
\begin{itemize}
\item[ (i)] The new Green's functions have the same properties with the old
ones. 

\item[(ii)] The above reshuffling does not change  the unique value of the
$S$-matrix, order by order in perturbation theory.

\end{itemize}

In what follows, we will show a ``mild'' version of uniqueness, namely that
the one-loop PT effective Green's functions are unique, provided that: 

\begin{itemize}

\item [(i)] The PT procedure can be generalized to higher orders in
perturbation theory, as described in \cite{JP&AP}. In particular, we assume
that effective GFP-independent Green's functions can be constructed, satisfying
the simple QED-like WI known from the one-loop explicit constructions, and that
the effective self-energies so constructed can be Dyson resummed. Regarding the
last point, the resummation algorithm proposed in \cite{JP&AP} not only is
inextricably connected to the fact that the PT self-energies do not shift the
position of the pole \cite{JP&AP}, but has already passed another non-trivial
consistency check \cite{KS}; still, one has not conclusively shown its
validity for the most general of cases.

\item [(ii)] The renormalization has been successfully carried out, giving
rise to UV finite effective PT Green's functions. This assumption is crucial,
and is the main reason why we characterize the uniqueness proved here as
``mild''. Things may be different if one attempts the aforementioned
reshuffling {\em before} renormalization, but this will not concern us in the
present work. 

\end{itemize} 

It is known \cite{PT} that the PT self-energy in QCD, $\widehat{\Pi}(q^2)$
(the lower and upper indices $T$ and $R$ are dropped for convenience),
captures the running of the coupling, exactly as happens in QED. To be
specific, setting 
\begin{equation}
\hat{d}_1(q^2)\ =\ \Big[\, q^2 + \widehat{\Pi}_1(q^2)\, \Big]^{-1}\, ,
\end{equation}
at one-loop, then the combination,
\begin{equation}
\widehat{D}_1(q^2)\ =\ g^2{\hat{d}}_1(q^2) \, ,
\label{D}
\end{equation}
obeys the 
following renormalization group equation (RGE):
\begin{equation}
\Big( \mu \frac{\partial}{\partial\mu}\ +\ g\beta_1 
                \frac{\partial}{\partial g}\Big)\widehat{D}_1(q^2)\ =\ 0\, ,
\label{RGE} 
\end{equation} 
where $\beta_1=-b_1\alpha_s/(4\pi)$. The reason for this is exactly the same
as in QED, namely the fact that the PT vertex and quark self energy satisfy an
Abelian, tree-level type Ward identity, {\em i.e.}, 
\begin{equation} 
q^{\mu}{\widehat{\Gamma}}_{\mu}\ =\
\widehat{\Sigma}(p+q)\, -\, \widehat{\Sigma}(p) 
\label{Z1Z2} 
\end{equation} 
or equivalently $\widehat{Z}_g=\widehat{Z}^{-1/2}_A$, where $\widehat{Z}_g$,
$\widehat{Z}_A$ are the gluon-field and strong-coupling-constant
renormalizations, respectively.

Let us now assume that the PT rearrangement, as described in \cite{JP&AP},
works to higher orders in perturbation theory. In particular, let us assume
that Eq.\ (\ref{RGE}) holds to all orders of perturbation, {\em i.e.}, for
\begin{equation}
\beta\ =\ -\, \Big[\, b_1\Big(\frac{\alpha_s}{4\pi}\Big)+
b_2\Big(\frac{\alpha_s}{4\pi}\Big)^2+\cdots+b_n
\Big(\frac{\alpha_s}{4\pi}\Big)^n+\cdots\, \Big]\, ,
\label{b12}
\end{equation}
and 
\begin{equation}
\widehat{\Pi}(q^2)\ =\ \widehat{\Pi}_1(q^2) +\widehat{\Pi}_2(q^2)
+ \cdots + \widehat{\Pi}_{n}(q^2) + \cdots\, ,
\label{Pi12}
\end{equation}  
where $\widehat{\Pi}_{n}$ are one-particle irreducible of $n$-loop order and
independent of the GFP. Note that the coefficients $b_{n}$ in Eq.(\ref{b12}) are
renormalization prescription dependent, for $n>2$.  The first three
coefficients for quark-less QCD are: 
\begin{equation}
b_1\ =\ \frac{11}{3}\, c_A\, , \qquad
b_2\ =\ \frac{34}{3}\, c^2_A\, ,\qquad
b_{3}\ =\ \frac{2857}{54}\, c^3_A\, ,
\end{equation}
and have been evaluated in Refs.\ \cite{HDP}, \cite{WEC} and \cite{TVZ},
respectively. The values of $b_{1}$ and $b_{2}$ quoted above are
renormalization scheme independent, whereas $b_{3}$ has been evaluated within
the minimal subtraction (MS) scheme \cite{GTH}. 

Substituting Eqs.\ (\ref{b12}) and (\ref{Pi12}) into Eq.\ (\ref{RGE}), and
equating powers of $g^2$, it is easy to obtain 
\begin{equation}
\label{RGQCD}
\mu \frac{\partial \widehat{\Pi}_n(q^2) }{\partial \mu}\ =
\ 2 \beta_{n}q^{2}\, +\, 2\sum_{k=1}^{n-1} 
(1-k)\beta_{n-k}\widehat{\Pi}_k(q^2)  \, ,
\end{equation} 
with $\beta_n=-b_n(a_s/4\pi)^n$. Notice that Eq.\ (\ref{RGQCD}) is identical
to the one obtained for the photon vacuum polarization in QED ~\cite{RR}. As
happens in the QED case, for $n=1,2$ the dependence of $\widehat{\Pi}_n $ on
the renormalization point $\mu$ is logarithmic, whereas for $n>2$, higher
powers of logarithms start appearing. 

Let us now assume that we were to change by hand the value of
$\widehat{\Pi}_1$, $\widehat{\Gamma}_1$ and $\widehat{B}_1$, in such a way as
to not change the value of the $S$-matrix at one loop. So, we make the
following replacements: 
\begin{eqnarray}
\widehat{\Pi}_1 & \rightarrow & \tilde{\Pi}_1\ \equiv\ \widehat{\Pi}_1 +
f_1\, , \nonumber\\ 
\widehat{\Gamma}_1 &\rightarrow & \tilde{\Gamma}_1
\ \equiv\ \widehat{\Gamma}_1 + u_1\, , \nonumber\\ 
 \widehat{B}_1 &\rightarrow &
{\tilde {B}}_1\ \equiv\ \widehat{B}_1 + h_1 \, ,
\label{Sub} 
\end{eqnarray}
where $f_1$, $u_1$ and $h_1$ are in principle arbitrary {\em functions} of
$q^2$, subject to the constraint
\begin{equation}
f_1+2q^2u_1+q^{4}h_1\ =\ 0  \, ,
\label{Constraint}
\end{equation}
which guarantees that the value of the $S$-matrix does not change at one loop,
after the substitution given in Eq.\ (\ref{Sub}). 

The functions $f_1$, $u_1$ and $h_1$ do not depend on the gauge fixing
parameter, and are UV and IR finite. Therefore, they do not
depend on the renormalization point $\mu$, {\em viz.}
\begin{displaymath} 
\frac{\partial {f}_1}{\partial\mu}\ =\ 
\frac{\partial {u}_1}{\partial\mu}\ =\ 
\frac{\partial {h}_1}{\partial\mu}=0\, . 
\end{displaymath}
In the case of QCD, the only physical choice for $f_1$ would be $f_1=Cq^2$,
where $C$ is a numerical constant, since the only available mass scale is
$q^2$. In other words, since $f$ does not depend on $\mu$, we cannot have
ratios of momenta $q^2/\mu^2$. At the same time, one does not want to use the
mass of the external fermions, since that would convert $\widehat{\Pi}_1$ to a
process-dependent quantity.  Moreover, the RGE in Eq.\ (\ref{RGQCD}) would then
be modified by the $\mu$ dependence of the running quark masses. For the sake
of argument, let us, however, assume that one uses a ``universal'' mass scale
$M_u$, such as the Planck mass, or some combination involving the sum of all
quark masses. So, $f_1$ may contain ratios of $q^2/M^2_u$. For example, $f_1$
could be of the form $f_1= q^2\exp (-q^2/M^2_u)$. However, it is important to
emphasize that $M_u$ should {\em not} depend on $\mu$, {\em i.e.}, $\partial
M_u/\partial\mu=0$. 
\begin{center}
\begin{picture}(340,200)(0,0)
\SetWidth{0.8}

\ArrowLine(0,150)(30,150)\Text(5,160)[r]{$q_1$}
\ArrowLine(30,150)(60,150)\Text(60,160)[r]{$q_1$}
\ArrowLine(0,40)(30,40)\Text(5,30)[r]{$q_2$}
\ArrowLine(30,40)(60,40)\Text(60,30)[r]{$q_2$}
\Gluon(30,130)(30,150){2}{2}
\Gluon(30,90)(30,100){2}{1}
\Gluon(30,40)(30,60){2}{2}
\BCirc(30,115){15}\Text(30,115)[c]{$\widehat{\Pi}_1$}
\BCirc(30,75){15}\Text(30,75)[c]{$\widehat{\Pi}_1$}
\Text(30,10)[c]{{\bf (a)}}

\Line(100,150)(130,150)
\Line(130,150)(160,150)
\Line(100,40)(130,40)
\Line(130,40)(160,40)
\Line(110,150)(130,120)
\Line(150,150)(130,120)\Text(130,140)[c]{$\widehat{\Gamma}_1$}
\Gluon(130,100)(130,120){2}{2}
\Gluon(130,40)(130,70){2}{3}
\BCirc(130,85){15}\Text(130,85)[c]{$\widehat{\Pi}_1$}
\Text(130,10)[c]{{\bf (b)}}

\Line(200,150)(230,150)
\Line(230,150)(260,150)
\Line(200,40)(230,40)
\Line(230,40)(260,40)
\Line(210,40)(230,70)
\Line(250,40)(230,70)\Text(230,50)[c]{$\widehat{\Gamma}_1$}
\Gluon(230,70)(230,100){2}{3}
\Gluon(230,130)(230,150){2}{2}
\BCirc(230,115){15}\Text(230,115)[c]{$\widehat{\Pi}_1$}
\Text(230,10)[c]{{\bf (c)}}

\Line(300,150)(330,150)
\Line(330,150)(360,150)
\Line(300,40)(330,40)
\Line(330,40)(360,40)
\Line(310,40)(330,70)
\Line(350,40)(330,70)\Text(330,50)[c]{$\widehat{\Gamma}_1$}
\Gluon(330,70)(330,120){2}{6}
\Line(310,150)(330,120)
\Line(350,150)(330,120)\Text(330,140)[c]{$\widehat{\Gamma}_1$}
\Text(330,10)[c]{{\bf (d)}}

\end{picture}\\
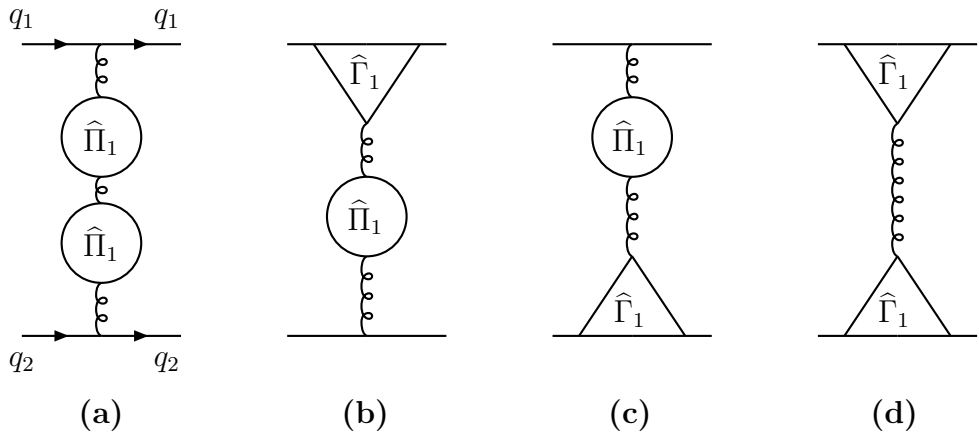
{\bf Fig.\ 8:} PT resummation at two loops in QCD.
    
\end{center}
Returning to the uniqueness issue, since the PT self-energies can be Dyson
summed \cite{JP&AP}, one should impose the same property on their new
counterparts. Therefore, following the method developed in \cite{JP&AP}, a
string of the form $\widehat{\Pi}_1\, (1/q^2)\, \widehat{\Pi}_1$ must be
converted to $\tilde{\Pi}_1\, (1/q^2)\, \tilde{\Pi}_1$. To accomplish this,
one must provide the appropriate combinations involving the functions $f_1$,
$u_1$, and $h_1$, just as we had to provide the missing pinch parts in going
from $\Pi_1\, (1/q^2)\, \Pi_1$ to $\widehat{\Pi}_1\, (1/q^2)\,
\widehat{\Pi}_1$ (see \cite{JP&AP}).  To see this in detail, we return to the
diagrams of Fig.\ 8, and assume that the PT rearrangement has already been
completed. So, now all bubbles and vertices in these graphs refer to the PT
objects. The relevant equations are
\begin{eqnarray}
\tilde{\Pi}_1\tilde{\Pi}_1&=&
(\widehat{\Pi}_1+f_1)(\widehat{\Pi}_1+f_1)\nonumber\\ 
&=& \widehat{\Pi}_1\widehat{\Pi}_1+2\widehat{\Pi}_1 f_1+ f^2_1\, ,
\\[0.3cm] 
\tilde{\Pi}_1\tilde{\Gamma}_1&=& (\widehat{\Pi}_1+f_1)
(\widehat{\Gamma}_1+ u_1)\nonumber\\ 
&=& \widehat{\Pi}_1\widehat{\Gamma}_1+
f_1\widehat{\Gamma}_1+ u_1\widehat{\Pi}_1+ f_1u_1\, ,\\[0.3cm] 
\tilde{\Gamma}_1\tilde{\Gamma}_1&=& (\widehat{\Gamma}_1+u_1)
(\widehat{\Gamma}_1+u_1) \nonumber\\ 
&=& 
\widehat{\Gamma}_1\widehat{\Gamma}_1+2u_1\widehat{\Gamma}_1+u_1^2\, .
\end{eqnarray}
Hereafter, the explicit $q^2$ dependence of the functions $\tilde{\Pi}$,
$\widehat{\Pi}$, $\tilde{\Gamma}$, etc., will not be displayed for brevity. 
Omitting a common factor of $(1/q^2)^3$, we obtain for the afore-mentioned
diagrams, 
\begin{equation}
\widehat{\Pi}_1\widehat{\Pi}_1+2q^2\widehat{\Pi}_1\widehat{\Gamma}_1+
q^{4}\widehat{\Gamma}_1\widehat{\Gamma}_1 \ =\
\tilde{\Pi}_1\tilde{\Pi}_1+2q^2\tilde{\Pi}_1
\tilde{\Gamma}_1+ q^{4}\tilde{\Gamma}_1\tilde{\Gamma}_1-R \, ,
\label{TiHat}
\end{equation}
with
\begin{equation}
R\ =\ (f_1+q^2u_1)\Big[ 2\widehat{\Pi}_1+2\widehat{\Gamma}_1+
(f_1+q^2u_1) \Big]\, .
\label{RQ}
\end{equation}
At one loop, the new effective charge $\tilde{D}_1$ satisfies the correct RGE.
In particular, since $\partial {f}/\partial\mu\, =\, 0$ by assumption, we have
that
\begin{equation}  
\mu \frac{\partial \tilde{\Pi}_1}{\partial \mu}\ =\ 
\mu \frac{\partial (\widehat{\Pi}_1+f_1)}{\partial \mu}= 2 \beta_1 q^{2} \, ,
\end{equation}
which is what Eq.\ (\ref{RGQCD}) yields for $n=1$.

According to the method in \cite{JP&AP}, the propagator-like parts of $R$ must
be allotted to $\Pi_2$. The second term in Eq.\ (\ref{RQ}) is
process-dependent, since it is proportional to $\widehat{\Gamma}_1$. This term
should be given to the two loop vertex or box graphs. In any case, as we will
see, this will make no difference in our analysis. But $\Pi_2$ has already
been converted into $\widehat{\Pi}_2$, because we assumed that the PT
procedure has been completed. Therefore, $\tilde{\Pi}_2$ must be defined as
follows: 
\begin{equation}
\tilde{\Pi}_2\ =\ \widehat{\Pi}_2+R_{2}^{p}\, ,
\label{TP2}
\end{equation}
where $R_{2}^{p}$ is the propagator-like part of $R_{2}$. After all
appropriate powers of $1/q^2$ have been restored, $R_{2}^{p}$ is given by 
\begin{equation}
R_{2}^{p}\ =\ \frac{2}{q^2}(f_1+q^2u_1)\widehat{\Pi}_1 + \dots,
\end{equation}
where the ellipses denote the optional inclusion of the third term in Eq.\
(\ref{RQ}), which is irrelevant for what follows, because it is 
$\mu$-independent.
 
It is now clear that $\tilde{\Pi}_2$ fails to satisfy the correct RGE, 
since its $\mu$-dependence is not in compliance with the result deduced
from Eq.\ (\ref{RGQCD}) for $n=2$. In particular, we have
\begin{eqnarray}
\mu \frac{\partial \tilde{\Pi}_2}{\partial \mu}
&=& \mu \frac{\partial}{\partial \mu} \Big[\widehat{\Pi}_2\, +\, 
\frac{2}{q^2}\, (f_1+q^2u_1)\widehat{\Pi}_1 \Big]  \nonumber \\ 
&=& 2 \beta_2 q^{2}\, +\, 4 \beta_1 (f_1+q^2u_1) \nonumber\\ 
&\neq & 2\beta_2 q^{2} \, .
\label{WRG}
\end{eqnarray}
So, in order to reconcile Dyson summation and the
correct RGE behaviour to the next order, we must impose the additional
constraint that
\begin{equation}
f_1+q^2u_1\ =\ 0\, .
\label{Constraint2}
\end{equation}
Combining this together with Eq.\ (\ref{Constraint}) we find that
$h_1=-u_1/q^4$. Thus, the entire expression for $R$ in Eq.\ (\ref{RQ})
vanishes, and Eq.\ (\ref{TiHat}) becomes 
\begin{equation}
\widehat{\Pi}_1\widehat{\Pi}_1+2q^2\widehat{\Pi}_1\widehat{\Gamma}_1+
q^{4}\widehat{\Gamma}_1\widehat{\Gamma}_1
= \tilde{\Pi}_1\tilde{\Pi}_1+2q^2\tilde{\Pi}_1
\tilde{\Gamma}_1+ q^{4}\tilde{\Gamma}_1\tilde{\Gamma}_1 \, .
\end{equation}

It appears at this point that we have succeeded in implementing the
substitution given in Eq.\ (\ref{Sub}), without compromising any of the PT
properties, at the seemingly modest expense of imposing on $f_1$ and $u_1$ the
additional constraint given in Eq.\ (\ref{Constraint2}).  However, as we will
see in a moment, Eq.\ (\ref{Constraint2}) is very crucial, because it actually
guarantees the uniqueness of our gauge-invariant resummation method
\cite{JP&AP}, at one-loop. 

To make this explicit, we proceed to the next order in perturbation theory. The
situation may be slightly more cumbersome calculationally, but the conceptual
issues are the same. By converting the old strings into new strings, we pick up
additional terms, which, when allotted to $\tilde{\Pi}_{3}$, these extra terms
will invalidate the RGE that $\tilde{\Pi}_{3}$ is expected to satisfy, 
{\em i.e.}, Eq.\ (\ref{RGQCD}) for $n=3$, unless a further constraint is
imposed on $f_1$. To determine that constraint, we focus on the three-loop
diagrams shown in Fig.\ 9. 

\begin{center}
\begin{picture}(340,500)(0,0)
\SetWidth{0.8}

\ArrowLine(0,490)(30,490)\Text(5,500)[r]{$q_1$}
\ArrowLine(30,490)(60,490)\Text(60,500)[r]{$q_1$}
\ArrowLine(0,340)(30,340)\Text(5,330)[r]{$q_2$}
\ArrowLine(30,340)(60,340)\Text(60,330)[r]{$q_2$}
\Gluon(30,470)(30,490){2}{2}
\Gluon(30,430)(30,440){2}{1}
\Gluon(30,390)(30,400){2}{1}
\Gluon(30,340)(30,360){2}{2}
\BCirc(30,455){15}\Text(30,455)[c]{$\widehat{\Pi}_1$}
\BCirc(30,415){15}\Text(30,415)[c]{$\widehat{\Pi}_1$}
\BCirc(30,375){15}\Text(30,375)[c]{$\widehat{\Pi}_1$}
\Text(30,310)[c]{{\bf (a)}}

\Line(100,490)(130,490)
\Line(130,490)(160,490)
\Line(100,340)(130,340)
\Line(130,340)(160,340)
\Gluon(130,460)(130,490){2}{3}
\Gluon(130,400)(130,430){2}{3}
\Gluon(130,340)(130,370){2}{3}
\BCirc(130,385){15}\Text(130,385)[c]{$\widehat{\Pi}_2$}
\BCirc(130,445){15}\Text(130,445)[c]{$\widehat{\Pi}_1$}
\Text(130,310)[c]{{\bf (b)}}

\Line(200,490)(230,490)
\Line(230,490)(260,490)
\Line(200,340)(230,340)
\Line(230,340)(260,340)
\Line(210,490)(230,460)
\Line(250,490)(230,460)\Text(230,480)[c]{$\widehat{\Gamma}_1$}
\Gluon(230,420)(230,460){2}{4}
\Gluon(230,340)(230,390){2}{5}
\BCirc(230,405){15}\Text(230,405)[c]{$\widehat{\Pi}_2$}
\Text(230,310)[c]{{\bf (c)}}

\Line(300,490)(330,490)
\Line(330,490)(360,490)
\Line(300,340)(330,340)
\Line(330,340)(360,340)
\Line(310,490)(330,460)
\Line(350,490)(330,460)\Text(330,480)[c]{$\widehat{\Gamma}_2$}
\Gluon(330,420)(330,460){2}{4}
\Gluon(330,340)(330,390){2}{5}
\BCirc(330,405){15}\Text(330,405)[c]{$\widehat{\Pi}_1$}
\Text(330,310)[c]{{\bf (d)}}

\Line(0,190)(30,190)
\Line(30,190)(60,190)
\Line(0,40)(30,40)
\Line(30,40)(60,40)
\Line(10,190)(30,160)
\Line(50,190)(30,160)\Text(30,180)[c]{$\widehat{\Gamma}_1$}
\Gluon(30,140)(30,160){2}{2}
\Gluon(30,90)(30,110){2}{2}
\Gluon(30,40)(30,60){2}{2}
\BCirc(30,125){15}\Text(30,125)[c]{$\widehat{\Pi}_1$}
\BCirc(30,75){15}\Text(30,75)[c]{$\widehat{\Pi}_1$}
\Text(30,10)[c]{{\bf (e)}}

\Line(100,190)(130,190)
\Line(130,190)(160,190)
\Line(100,40)(130,40)
\Line(130,40)(160,40)
\Line(110,40)(130,70)
\Line(150,40)(130,70)\Text(130,50)[c]{$\widehat{\Gamma}_1$}
\Gluon(130,70)(130,100){2}{3}
\BCirc(130,115){15}\Text(130,115)[c]{$\widehat{\Pi}_1$}
\Gluon(130,130)(130,160){2}{3}
\Line(110,190)(130,160)
\Line(150,190)(130,160)\Text(130,180)[c]{$\widehat{\Gamma}_1$}
\Text(130,10)[c]{{\bf (f)}}

\Line(200,190)(230,190)
\Line(230,190)(260,190)
\Line(200,40)(230,40)
\Line(230,40)(260,40)
\Line(210,40)(230,70)
\Line(250,40)(230,70)\Text(230,50)[c]{$\widehat{\Gamma}_2$}
\Gluon(230,70)(230,160){2}{9}
\Line(210,190)(230,160)
\Line(250,190)(230,160)\Text(230,180)[c]{$\widehat{\Gamma}_1$}
\Text(230,10)[c]{{\bf (g)}}

\end{picture}\\
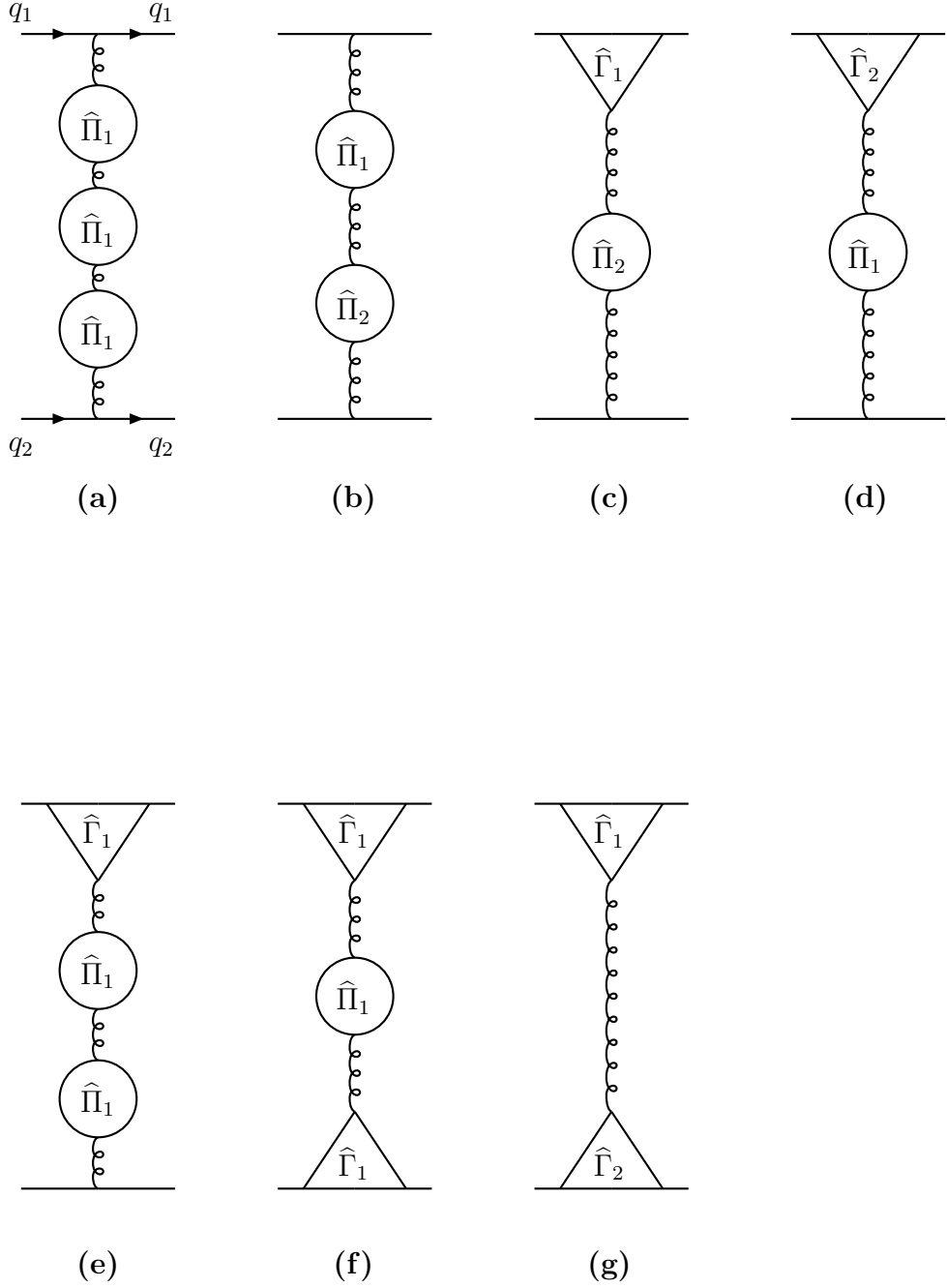
{\bf Fig.\ 9:} PT resummation at three loops in QCD.
    
\end{center}

Again, in order to be as general as possible, we assume that one can reshuffle
the second order PT Green's functions, without affecting the value of the
$S$-matrix to that order. In other words, we allow the additional
substitutions 
\begin{eqnarray}
\widehat{\Pi}_2 &\rightarrow& \tilde{\Pi}_2\ \equiv\ \widehat{\Pi}_2 +
f_2\, , \nonumber\\ 
\widehat{\Gamma}_2 &\rightarrow& \tilde{\Gamma}_2\
\equiv\ \widehat{\Gamma}_2 + u_2\, , \nonumber\\ 
\widehat{B}_2 &\rightarrow&
\tilde{B}_2\ \equiv\ \widehat{B}_2 + h_2 \, ,
\label{Sub2L} 
\end{eqnarray}
with 
\begin{equation}
f_2+2q^2u_2+q^{4}h_2\ =\ 0 \, .
\label{Constraint2L}
\end{equation}
Of course the proof becomes easier if we assume $f_2=u_2=h_2=0$, but we do not
have to.  We will need the following algebraic relations:
\begin{eqnarray}
\tilde{\Pi}_1^{3}&=& {(\widehat{\Pi}_1+f_1)}^{3} \nonumber\\
&=& \widehat{\Pi}_1^{3} + 3\widehat{\Pi}_1^2f_1 + 3
\widehat{\Pi}_1 f_1^2+ f_1^{3}\, ,\\[0.3cm] 
\tilde{\Pi}_1\tilde{\Pi}_2 &=&
(\widehat{\Pi}_1+f_1)(\widehat{\Pi}_2+f_2) \nonumber\\
&=& \widehat{\Pi}_1\widehat{\Pi}_2 + \widehat{\Pi}_1 f_2 + 
\widehat{\Pi}_2f_1+ f_1f_2\, ,\\[0.3cm]
\tilde{\Pi}_1{\tilde {\Gamma}}_2 &=& (\widehat{\Pi}_1+f_1)(
\widehat{\Gamma}_2+u_2) \nonumber\\ 
&=& \widehat{\Pi}_1\widehat{\Gamma}_2 +
\widehat{\Pi}_1u_2+ f_1\widehat{\Gamma}_2+ f_1u_2\, ,\\[0.3cm] 
\tilde{\Pi}_2\tilde{\Gamma}_1 &=& (\widehat{\Pi}_2+f_2)(
\widehat{\Gamma}_1+u_1) \nonumber\\ 
&=& \widehat{\Pi}_2\widehat{\Gamma}_1 +
\widehat{\Pi}_2u_1+ f_2\widehat{\Gamma}_1+ f_2u_1\, ,\\[0.3cm]
\tilde{\Pi}_1^2\tilde{\Gamma}_1 
&=& (\widehat{\Pi}_1+f_1)^2(\widehat{\Gamma}_1+u_1) \nonumber\\ 
&=& 
\widehat{\Pi}_1^2\widehat{\Gamma}_1+ u_1\widehat{\Pi}_1^2+ 2
f_1u_1\widehat{\Pi}_1 + 2 f_1\widehat{\Pi}_1\widehat{\Gamma}_1 +
f_1^2\widehat{\Gamma}_1 + f_1^{2} u_1 \, , \\[0.3cm]
\tilde{\Pi}_1\tilde{\Gamma}_1^2 &=&  
(\widehat{\Pi}_1+f_1)(\widehat{\Gamma}_1+u_1)^2 \nonumber\\ 
&=& \widehat{\Pi}_1\widehat{\Gamma}_1^2+u_1^2
\widehat{\Pi}_1+2u_1\widehat{\Pi}_1\widehat{\Gamma}_1 + 
f_1\widehat{\Gamma}_1^2 + 2 f_1u_1\widehat{\Gamma}_1 + f_1u_1^2\, ,\\[0.3cm] 
\tilde{\Gamma}_1\tilde{\Gamma}_2 &=& 
(\widehat{\Gamma}_1+u_1)(\widehat{\Gamma}_2+u_2) \nonumber\\ 
&=&\widehat{\Gamma}_1 \widehat{\Gamma}_2 + u_2 \widehat{\Gamma}_1 +
u_1\widehat{\Gamma}_2 + u_1u_2\, .
\end{eqnarray}
Using the above formulas, the crucial constraint of Eq.\ (\ref{Constraint2}),
and remembering that the graphs of the Figs. 9(b)--(e) and 9(g) must be
multiplied by a factor of 2, which takes account of the symmetric (mirror
image) graphs, we have that the original set of graphs, call $\hat{\cal A}$
(we factor out a factor $(1/q^2)^{4}$ ) 
\begin{equation}
\hat{\cal A}\ =\
\widehat{\Pi}_1^{3}+ 2 q^2 ( \widehat{\Pi}_1\widehat{\Pi}_2 + 
\widehat{\Pi}_1^2 \widehat{\Gamma}_1 ) + q^{4} (2\widehat{\Pi}_1
\widehat{\Gamma}_2 + 2\widehat{\Pi}_2\widehat{\Gamma}_1 + 
\widehat{\Pi}_1 \widehat{\Gamma}_1^2 ) + 
2q^{6}\widehat{\Gamma}_1 \widehat{\Gamma}_2
\label{hatcalA} 
\end{equation}
and the new one, $\tilde{\cal A}$ say, which is obtained by replacing all
``hatted'' quantities in Eq.\ (\ref{hatcalA}) by ``tilded'' ones, are related
by
\begin{equation}
\hat{\cal A}\ =\ \tilde{\cal A} - R_{3} \, ,
\end{equation} 
where $R_{3}$ is given by
\begin{eqnarray}
R_{3} &=& f_1 \widehat{\Pi}_1^2 + 2 q^2(f_2+q^2u_2)\widehat{\Pi}_1 + 
2q^2f_1\widehat{\Pi}_1\widehat{\Gamma}_1 
 + q^{4}f_1 \widehat{\Gamma}_1^2\nonumber\\
&&+ 2 q^{4}(f_2+q^2u_2)\widehat{\Gamma}_1 \, .
\label{TotalR}    
\end{eqnarray}
Clearly, the first two terms in Eq.\ (\ref{TotalR}) must be allotted to
$\widehat{\Pi}_3$, thus converting it to $\tilde{\Pi}_{3}$. The rest of the
terms cannot be absorbed by $\tilde{\Pi}_{3}$, since they are explicitly
process-dependent, because they contain $\widehat{\Gamma}_1$. Therefore, the
remaining terms must be distributed among the two-loop vertex and/or box graphs.
So, after all powers of $1/q^2$ are restored, the propagator-like part
$R_{3}^{p}$ of $R_{3}$ reads 
\begin{equation} 
R_{3}^{p}\ =\ \frac{f_1}{q^{4}}\, \widehat{\Pi}_1^2\, 
+\, \frac{2}{q^2}\, (f_2+q^2u_2)\widehat{\Pi}_1\, , 
\label{R3P}
\end{equation}
and so
\begin{equation}
\tilde{\Pi}_3\ =\ \widehat{\Pi}_3\, +\, R_{3}^{p}\, .
\label{Tild3} 
\end{equation}

It is now important to observe that, because of the particular structure of
$R_{3}^{p}$, the RGE satisfied by $\tilde{\Pi}_{3}$ will be modified. Indeed,
from Eq.\ (\ref{RGQCD}), we derive for $n=3$
\begin{equation}
\mu \frac{\partial \widehat{\Pi}_3 }{\partial \mu}\ =
\ 2 \beta_{3}q^{2}\, -\, 2\beta_1\widehat{\Pi}_2 \ 
\label{RGEM}
\end{equation}
and after the substitution $\widehat{\Pi}_i \rightarrow \tilde{\Pi}_{i}$, 
we must have 
\begin{equation}
\mu \frac{\partial \tilde{\Pi}_3 }{\partial \mu}\ =
\ 2 \beta_{3}q^{2}\, -\, 2\beta_1\tilde{\Pi}_2 \, .
\label{RGEH}
\end{equation}
Subtracting the two last equations by parts, we obtain
\begin{eqnarray}
\mu \frac{\partial}{\partial \mu}\, (\tilde{\Pi}_3 - \widehat{\Pi}_3) &=&
- 2 \beta_{1}(\tilde{\Pi}_2 -\widehat{\Pi}_{2}) \nonumber \\ 
&=& -2 \beta_{1}f_{2}\, .
\label{RGED}
\end{eqnarray}
Instead, from Eqs.\ (\ref{R3P}) and (\ref{Tild3}), we find
\begin{eqnarray}
\mu \frac{\partial }{\partial \mu}\, (\tilde{\Pi}_3 - \widehat{\Pi}_3) &=&
\mu \frac{\partial R_{3}^{p}}{\partial \mu} \nonumber \\
&=& \frac{4f_1}{q^{2}}\, \beta_{1}\widehat{\Pi}_1\,  +\, 
4\beta_{1}(f_2+q^2u_2)\, .
\label{RGER}
\end{eqnarray}  
Given the fact that $\widehat{\Pi}_1$ depends explicitly on $\mu$, in order to
reconcile Eqs.\ (\ref{RGED}) and (\ref{RGER}) one must necessarily choose
$f_1=0$. Thus, the only possible solution for the set of substitutions
described in Eq.\ (\ref{Sub}) is the {\em trivial} one, {\em i.e.},
$f_1=u_1=h_1=0$, which proves the uniqueness of the PT resummation approach to
one-loop, after renormalization. 

After setting $f_1=0$, we must impose the additional constraint $3f_{2}+2q^2
u_2=0$, in order that Eqs.\ (\ref{RGED}) and (\ref{RGER}) become equal.
Evidently, the same arguments presented above must be repeated to the next
order, which will finally determine the value of $f_{2}$; we will not pursue
this issue any further here. Instead, we add some further clarifications
regarding the assumptions made in the previous proof of the one-loop
uniqueness of the PT resummation formalism. As emphasized at the beginning of
this section, we assume that the PT can be extended to higher orders, giving
rise to effective Green's function with all the characteristics known from the
explicit one-loop analysis. We further assume that the renormalization
programme has been carried out to all orders. Thus, all ``hatted'' Green's
functions appearing are UV finite. So far, the renormalization scheme chosen 
has been left unspecified. Because of Eq.\ (\ref{RGQCD}), the effect of 
adopting different renormalization-scheme choices will be to modify the 
values of $b_{n}$, for $n>2$. However, within a specific renormalization 
scheme, the values of $b_{n}$ are fixed, and this is what we have implicitly 
assumed. 

The resummation formalism discussed for the case of Yang-Mills theories
such as QCD can equally carry over to SSB models such as the SM.  In
the SM, $W$ and $Z$ bosons are considered to be unstable gauge particles.
In the case of the $W$ boson, a RGE similar to Eq.\ (\ref{RGQCD}) will
hold for the leading logarithmic part of the transverse $W$-boson 
self-energy. Again, one can form the RGE invariant combination involving the
$W$-boson Green's function 
\begin{displaymath}
g^2_w\, \Big[\, q^2\, +\, \widehat{\Pi}^W_T(q^2)\, \Big]^{-1}\, .
\end{displaymath}
Analogously with Eq.\ (\ref{Z1Z2}), one can derive a similar relation between
the weak-coupling-constant renormalization $\widehat{Z}_{g_w}$ and the
wave-function renormalization of the $W$ boson $\widehat{Z}_W$, 
{\em i.e.}, $\widehat{Z}_{g_w}=\widehat{Z}_W^{-1/2}$. Hence, one 
can show the uniqueness of this expression by following a line of arguments
similar to the case of QCD. Furthermore, possible modifications of the
longitudinal part of the $W$-boson self-energy, $\widehat{\Pi}_L^W$, will
result in direct violations of the tree-level WIs, which govern the gauge
invariance of the classical action.

\setcounter{equation}{0}
\section{Conclusions}
\indent

We have presented a formalism for resummation of off-shell two-point
correlation functions, which relies entirely on arguments of analyticity,
unitarity, gauge invariance and multiplicative renormalization. In addition,
several crucial aspects of the GFP-independent resummation approach
presented in \cite{JP&AP} have been clarified. Specifically, we have shown
that unitarity requires the absence of unphysical thresholds for the resummed
Green's functions at the quantum loop level. Within the PT resummation
approach this property is satisfied, since the effective gauge-invariant
Green's functions are directly derived from $S$-matrix elements, with the only
additional input the use of elementary tree-level WIs and analyticity. 

This is, however, not true in other approaches. For instance, we have explicitly
shown that $\xi_Q$-dependent unphysical thresholds appear in the BFG, even
though the Green's functions obey the same tree-level WIs as the PT Green's
functions. For the very specific value of $\xi_Q=1$, the results of BFG and PT
coincide to one-loop, as this is the only gauge that avoids unphysical
propagator poles. The situation may change in higher orders. Furthermore, we
have found that the BFG Green's functions can be decomposed into two parts,
one containing only physical poles and one containing $\xi_Q$-dependent
unphysical thresholds, which {\em separately} satisfy the same WIs as the
total BFG Green's functions. 

Furthermore, we have addressed issues of gauge invariance by resorting to the
BRS symmetries at the one-loop quantum level. We have explicitly demonstrated
that the PT two-point correlation function may be obtained from its absorptive
part through a DR. The absorptive part of the PT Green's
functions can equally well be calculated from the optical relation of the
anti-hermitian part of the transition amplitude.  As a result of this, we have
also been able to identify the pinching parts of the PT algorithm, as those
terms that quantify the deviation from the intrinsic BRS symmetries. Most
importantly, we have been able to show how gauge invariance is restored,
within the PT framework, by reinforcing BRS symmetries inside the quantum loops. 

In Section 9, we have examined the issue of ``uniqueness'' of the
gauge-invariant resummation approach proposed in \cite{JP&AP}. In the context
of QCD, we have focused on the most basic RGE invariant quantity involving the
PT two-point correlation function, namely the effective (running) strong
coupling. By means of a three-loop analysis, we have shown that, at one-loop,
the PT resummation method gives rise to unique results. We have also briefly
outlined how these considerations can be naturally extended to spontaneously
broken gauge theories. 

Considering the fact that all the basic field-theoretical requirements imposed
thus far are preserved within the PT resummation approach that was introduced
in \cite{JP&AP} and was further analysed in the present paper, one might be
tempted to argue that some deeper underlying principle is in effect, which has
yet to be discovered. Here we wish to point out two possibly
relevant directions in such a quest. 
First, there is a interesting recent result of ``stringy'' origin
\cite{VMLRM}, which seems to single out the one-loop BFG Green's functions for
the special value of $\xi_{Q}=1$, which are, of course, identical to the PT
Green's functions. This observation makes the question of whether the
correspondence between the PT and the BFG at $\xi_{Q}=1$ persists beyond one
loop even more pressing. Second, one should investigate possible connections
between the PT and the Vilkovisky-DeWitt formalism \cite{VDW}. In particular,
the gauge invariant and GFP-independent Green's functions obtained from the
Vilkovisky-DeWitt effective action must be compared with their PT
counterparts, establishing the origin and the physical significance of any
possible difference between them.\\[0.7cm] 
{\bf Acknowledgements.} The authors thank J.D. Bjorken, H.-M. Chan,
J.M. Cornwall, G. Gounaris, E. Kiritsis, C. Kounnas, E. de Rafael,
G. Veneziano, and J. Watson for several useful discussions.


\begin{thebibliography}{99}
\bibitem{RGE} In the framework of the $S$-matrix perturbation theory, the
evaluation of an $S$-matrix element or a gauge-invariant operator at a given
order of loop expansion is not unique in general, in the sense that the
analytic result depends on the renormalization prescription used to remove the
UV divergences. Of course, the summation of all infinite
perturbative contributions should formally yield a unique result independent of
the choice of renormalization. Furthermore, to a given order of perturbation
theory, one can invoke the renormalization group equation (RGE) in order to 
show that this uniqueness of the final expression gets spoiled by terms which
are of higher order in the coupling constants. The latter notion of RGE
invariance is to be adopted throughout this paper.
\end{thebibliography}
\end{document}